\documentclass{ptephy}%%%%where ptephy is the template name

\usepackage{amsmath} % for dealing with mathematics,
\usepackage{amsthm} % for dealing with theorem environments,
\usepackage{graphics} % for dealing with figures.

\usepackage{bm}
\usepackage{color}
\usepackage{ulem}

\renewcommand{\sout}{\bgroup \color[rgb]{1,0,0} \ULdepth=-.5ex \ULset}
\newcommand{\PsfigII}[2]{\includegraphics[scale=#1]{#2}}

\def\mev{\text{ MeV}}
\def\gev{\text{ GeV}}

\def\naive{na\"{i}ve }
\def\Kaellen{K\"{a}llen }
\def\Schr{Schr\"{o}dinger }

\begin{document}

\title{Comprehensive analysis of the wave function of a hadronic resonance
  and its compositeness}

\author{\name{Takayasu Sekihara}{1,\ast}
  \name{Tetsuo Hyodo}{2}, 
  and \name{Daisuke Jido}{3}}
%%%%%%%%%%% The \name command should be used as \name{Insert author name here}{Insert affiliation number here}
%%%%% Please use \thanks for contributed author details

%%%%%%%%%%% The \affil command should be used as \affil{Insert affiliation number here}{Insert author address here}
\address{
  \affil{1}{Research Center for Nuclear Physics (RCNP), Osaka University,
    Ibaraki, Osaka, 567-0047, Japan} %
  \affil{2}{Yukawa Institute for Theoretical Physics, 
    Kyoto University, Kyoto 606-8502, Japan} %
  \affil{3}{Department of Physics, Tokyo Metropolitan University,
    Hachioji 192-0397, Japan} %
  \email{sekihara@rcnp.osaka-u.ac.jp} %
}
%\date{\today}% Should be deleted before submission!!!

\begin{abstract}%
  We develop a theoretical framework to investigate the two-body
  composite structure of a resonance as well as a bound state from its
  wave function.  For this purpose, we introduce both one-body bare
  states and two-body scattering states, and define the compositeness
  as a fraction of the contribution of the two-body wave function to
  the normalization of the total wave function.  Writing down
  explicitly the wave function for a resonance state obtained with a
  general separable interaction, we formulate the compositeness in
  terms of the position of the resonance pole, the residue of the
  scattering amplitude at the pole and the derivative of the Green
  function of the free two-body scattering system.  At the same time,
  our formulation provides the elementariness expressed with the
  resonance properties and the two-body effective interaction, and
  confirms the sum rule showing that the summation of the
  compositeness and elementariness gives unity.  In this formulation
  the Weinberg's relation for the scattering length and effective
  range can be derived in the weak binding limit.  The extension to
  the resonance states is performed with the Gamow vector, and a
  relativistic formulation is also established.
  As its applications, we study the compositeness of the $\Lambda
  (1405)$ resonance and the light scalar and vector mesons described
  with refined amplitudes in coupled-channel models with interactions
  up to the next to leading order in chiral perturbation theory.  We
  find that $\Lambda (1405)$ and $f_{0}(980)$ are dominated by the
  $\bar{K} N$ and $K \bar{K}$ composite states, respectively, while
  the vector mesons $\rho (770)$ and $K^{\ast} (892)$ are elementary.
  We also briefly discuss the compositeness of $N (1535)$ and $\Lambda
  (1670)$ obtained in a leading-order chiral unitary approach. 
\end{abstract}

%\subjectindex{xxxx, xxx}

\maketitle

%\tableofcontents

\section{Introduction}

% exotic structures of hadrons
In hadron physics, the internal structure of an individual hadron is
one of the most important subjects.  Traditionally, the excellent
successes of constituent quark models lead us to the interpretation
that baryons consist of three quarks ($qqq$) and mesons of a
quark-antiquark pair ($q\bar{q}$)~\cite{Agashe:2014kda}. At the same
time, however, there are experimental indications that some hadrons do
not fit into the classification suggested by constituent quark models.
One of the classical examples is the hyperon resonance $\Lambda
(1405)$, which has an anomalously light mass among the negative parity
baryons. In addition, the lightest scalar mesons [$f_{0}(500) =
\sigma$, $K_{0}^{\ast}(800) = \kappa$, $f_{0}(980)$, and $a_{0}(980)$]
exhibit inverted spectrum from the \naive expectation with the
$q\bar{q}$ configuration. These observations motivate us to consider
more exotic structure of hadrons, such as hadronic molecules and
multiquarks~\cite{Dalitz:1960du, Dalitz:1967fp, Jaffe:1976ig,
  Jaffe:1976ih, Weinstein:1982gc, Weinstein:1983gd}.

% experimental and theoretical activities
It is encouraging that there have been experimental reports on the
candidates of manifestly exotic hadrons such as charged
quarkonium-like states by Belle collaboration~\cite{Belle:2011aa}.
Moreover, the LEPS collaboration observed the ``$\Theta ^{+}$
signal''~\cite{Nakano:2003qx,Nakano:2008ee}, but its interpretation is
still controversial~\cite{MartinezTorres:2010zzb, Torres:2010jh}.  The
accumulation of the observations of unconventional states in the heavy
quark sector reinforces the existence of hadrons with exotic
structure~\cite{Swanson:2006st, Brambilla:2010cs}.  In fact, recent
detailed analyses of $\Lambda (1405)$ in various
reactions~\cite{Niiyama:2008rt, Moriya:2013eb, Lu:2013nza,
  Agakishiev:2012xk} and of the $a_{0}^{0}(980)$-$f_{0}(980)$ mixing
in $J/\psi$ decay~\cite{Ablikim:2010aa} are providing some clues for
unusual structure of these hadrons. The exotic structure is also
investigated by analyzing the theoretical models; the meson-baryon
components of $\Lambda(1405)$ by using the natural renormalization
scheme~\cite{Hyodo:2008xr}, the $N_{c}$ scaling behaviors of scalar
and vector mesons~\cite{Pelaez:2003dy, Pelaez:2004xp} and of
$\Lambda(1405)$~\cite{Hyodo:2007np, Roca:2008kr}, spatial size of
$\Lambda(1405)$~\cite{Sekihara:2008qk, Sekihara:2010uz,
  Sekihara:2012xp}, $\sigma$ meson~\cite{Albaladejo:2012te}, and
$f_{0}(980)$~\cite{Sekihara:2012xp}, the nature of the $\sigma$ meson
from the partial restoration of chiral symmetry~\cite{Hyodo:2010jp},
and the structure of $\sigma$ and $\rho (770)$ mesons studied by their
Regge trajectories~\cite{Londergan:2013dza}.  The possibilities to
extract the hadron structure from the production yield in relativistic
heavy ion collisions~\cite{Cho:2010db, Cho:2011ew} and from the
high-energy exclusive productions~\cite{Kawamura:2013iia,
  Kawamura:2013wfa} are also suggested.

% hadronic molecule structure
Among various exotic structures, hadronic molecular configurations are
of special interest. These states are composed of two (or more)
constituent hadrons by strong interaction between them without losing
the character of constituent hadrons, in a similar way with the atomic
nuclei as bound states of nucleons. The $\bar{K}N$ quasi-bound picture
for $\Lambda (1405)$ is one of the examples. In contrast to the quark
degrees of freedom, the masses and interactions of hadrons are defined
independently of the renormalization scheme of QCD, because hadrons
are color singlet states. This fact implies that the structure of
hadrons may be adequately defined in terms of the hadronic degrees of
freedom. This viewpoint originates in the investigations of the
elementary or composite nature of particles in terms of the field
renormalization constant~\cite{Salam:1962ap, Weinberg:1962hj,
  Ezawa:1963zz}. Indeed, it is shown in this approach that the
deuteron is dominated by the loosely bound proton-neutron
component~\cite{Weinberg:1965zz}. The study of the structure of
hadrons from the field renormalization constant have been further
developed in Refs.~\cite{Baru:2003qq, Hanhart:2007cm, Hanhart:2011jz,
  Hyodo:2011qc, Aceti:2012dd, Xiao:2012vv, Hyodo:2013iga,
  Hyodo:2013nka, Sekihara:2013sma, Aceti:2014ala, Aceti:2014wka,
  Nagahiro:2014mba, Sekihara:2014qxa}.

% summary of the present work
Motivated by these observations, in this study we develop a framework
to investigate hadronic two-body components inside a hadron by
analyzing comprehensively the wave function of a resonance state.  For
this purpose, we explicitly introduce one-body bare states in addition
to the two-body components so as to form a complete set within them
and to measure the elementary and composite contributions.  The
one-body component has not been taken into account in the preceding
studies on wave functions (see Refs.~\cite{Aceti:2012dd,
  Gamermann:2009uq, YamagataSekihara:2010pj}).  For the resonance
state we employ the Gamow vector~\cite{Gamow:1928zz}, which ensures a
finite normalization of the resonance wave function.  The wave
function from a relativistically covariant wave equation is also
discussed.  Making a good use of a general separable interaction, we
analytically solve the wave equations.

In the present formulation, the compositeness and elementariness are
respectively defined as the fractions of the contributions from the
two-body scattering states and one-body bare states to the
normalization of the total wave function.  They are further expressed
with the quantities in the scattering equation with a general
separable interaction.  As a consequence, the compositeness can be
written in terms of the residue of the scattering amplitude at the
pole position, i.e., the coupling constant, and the derivative of the
Green function of the free two-body scattering system at the pole.
This means that the compositeness can be obtained solely with the pole
position of the resonance and the residue at the pole but without
knowing the details of the two-body effective interaction.  On the
other hand, the elementariness is obtained with the residue of the
scattering amplitude, the Green function and the derivative of the
two-body effective interaction at the pole.  It is an interesting
finding that with this expression we are allowed to interpret the
elementariness as the contributions coming from one-body bare states
and implicit two-body channels which do not appear as explicit degrees
of freedom but are effectively taken into account for the two-body
interaction in the practical model space.  Through the discussion on
the multiple bare states, we show that our formulation of the
compositeness and elementariness can be applied to any separable
interactions with arbitrary energy dependence.  Based on this
foundation, as applications we evaluate the compositeness of hadronic
resonances, such as $\Lambda (1405)$, the light scalar mesons and
vector mesons described in the chiral coupled-channel approach with
the next-to-leading order interactions so as to discuss their internal
structure from the viewpoint of hadronic two-body components.

% paper organization
This paper is organized as follows. In Sec.~\ref{sec:2}, we formulate
the compositeness and elementariness of a physical particle state in
terms of its wave function, and show their connection to the physical
quantities in scattering equation.  We first consider a two-body bound
state in the nonrelativistic framework, and later extend the
formulation to a resonance state in a relativistic covariant form with
the Gamow vector.  In Sec.~\ref{sec:3} numerical results for the
applications to physical resonances are presented. Section~\ref{sec:4}
is devoted to drawing the conclusion of this study.

\section{Compositeness and elementariness from wave functions}
\label{sec:2}

In this section, we define the compositeness (and simultaneously
elementariness) of a particle state, i.e., a stable bound state or a
unstable resonance, using its wave function and link the compositeness
to the physical quantities in scattering equation.  For this purpose,
we consider two-body scattering states\footnote{We note that the
  two-body wave functions are given by the asymptotic states of the
  system. In the application to QCD, the basis should be spanned by
  the hadronic degrees of freedom. The compositeness in terms of
  quarks cannot be defined in this approach.} coupled with each other
and one-body bare states. The one-body bare states have not been
introduced in the studies of wave functions before, and the
introduction of the one-body bare states makes it clear to implement
the elementariness into the formulation. To solve the scattering
equation analytically, we make use of the separable type of
interaction. We will concentrate on an s-wave scattering system, and
thus the two-body wave function and the form factors are assumed to be
spherical.

In Sec.~\ref{sec:2-1} we consider a bound state\footnote{In general,
  there can be several bound states in the system.  In such a case, we
  just focus on a bound state out of these bound states. Nothing
  changes in the following discussion. } in two-body scattering.  We
first introduce a one-body bare state and a single scattering channel,
and give the expressions of the compositeness and the elementariness
in terms of the wave function of the bound state.  In
Sec.~\ref{sec:2-2} we extend the discussion to a system with multiple
bare states and coupled scattering channels, in order to clarify
further the meaning of the compositeness and elementariness obtained
in Sec.~\ref{sec:2-1}.  Here we also discuss a way to introduce
general energy dependent interaction into the formulation. In
Sec.~\ref{sec:2-3} we consider the weak binding limit to derive the
Weinberg's relation for the scattering length and the effective
range~\cite{Weinberg:1965zz}. Generalization to resonance states is
discussed in Sec.~\ref{sec:2-4}. Finally we give a relativistic
covariant formulation in Sec.~\ref{sec:2-5}.

\subsection{Bound state in the nonrelativistic scattering}
\label{sec:2-1}

We consider a two-body scattering system in which there exists a
discrete energy level below the scattering threshold energy.  We call
this energy level bound state since it is located below the two-body
scattering threshold.  We do not assume the origin and structure of
the bound state at all.  We take the rest frame of the center-of-mass
motion, namely two scattering particles have equal and opposite
momentum and the bound state is at rest with zero momentum. The system
in this frame is described by Hamiltonian $\hat{H}$ which consists of
the free part $\hat{H}_0$ and the interaction term $\hat{V}$
\begin{equation}
   \hat{H} = \hat{H}_{0} + \hat{V} .
\end{equation}
We assume that the free Hamiltonian has continuum eigenstates $|
\bm{q} \rangle$ for the scattering state and one discrete state $|
\psi _{0} \rangle$ for the one-body bare state.  The eigenvalues of
the Hamiltonian are set to be
\begin{align}
\hat{H}_{0} | \bm{q} \rangle 
&= \left ( M^{\rm th} + \frac{q^{2}}{2 \mu} \right ) | \bm{q} \rangle , 
\quad 
\langle \bm{q} | \hat{H}_{0} 
= \left ( M^{\rm th} + \frac{q^{2}}{2 \mu} \right ) \langle \bm{q} | , 
\\
\hat{H}_{0} | \psi _{0} \rangle 
&= M_{0} | \psi _{0} \rangle , 
\quad 
\langle \psi _{0} | \hat{H}_{0} = M_{0} \langle \psi _{0} | , 
\end{align}
where $\mu$ is the reduced mass of the two-body system, $M_{0}$ is the
mass of the bare state, and $q \equiv |\bm{q}|$.  We include the sum
of the scattering particle masses, $M^{\rm th}$, which is just the
scattering threshold energy, into the definition of the eigenenergy
for later convenience.  These eigenstates are normalized as
\begin{equation}
\langle \bm{q}^{\prime} | \bm{q} \rangle 
= (2 \pi )^{3} \delta ^{3} (\bm{q}^{\prime} - \bm{q}) , \quad
\langle
\psi _{0}| \psi _{0} \rangle
=1 , \quad
\langle \psi _{0}| \bm{q} \rangle 
= \langle \bm{q} | \psi
_{0} \rangle =0 .
\end{equation}
These states form the complete set of the free Hamiltonian, and thus
we can decompose unity in the following way
\begin{equation}
1 = | \psi _{0} \rangle \langle \psi _{0} | 
+ \int \frac{d^{3} q}{(2 \pi )^{3}} 
| \bm{q} \rangle  \langle \bm{q} | .
\label{eq:unityQM}
\end{equation}

The bound state is realized as an eigenstate of the full Hamiltonian: 
\begin{equation}
\hat{H} | \psi \rangle = M_{\rm B} | \psi \rangle , 
\quad 
\langle \psi | \hat{H} = M_{\rm B} \langle \psi | ,
\label{eq:Schroedinger}
\end{equation}
where $M_{\rm B}$ is the mass of the bound state. The bound state wave
function is normalized as
\begin{equation}
\langle \psi | \psi \rangle = 1 .
\label{eq:normalization}
\end{equation}

We take the matrix element of Eq.~\eqref{eq:unityQM} in terms of the
bound state $| \psi \rangle$:
\begin{equation}
1 = \langle \psi | \psi _{0} \rangle 
\langle \psi _{0} | \psi \rangle 
+ \int \frac{d^{3} q}{( 2 \pi )^{3}} 
\langle \psi | \bm{q} \rangle 
\langle \bm{q} | \psi \rangle .
\end{equation}
The first term of the right-hand side is the probability of finding
the bare state in the bound state and also corresponds to the field
renormalization constant in the field theory.  Thus, we call this
quantity elementariness $Z$:
\begin{equation}
Z \equiv \langle \psi | \psi _{0} \rangle 
\langle \psi _{0} | \psi \rangle . 
\end{equation}
Because $\langle \psi | \psi _{0} \rangle =\langle \psi _{0} | \psi
\rangle^{*}$, $Z$ is always real and nonnegative. The second term, on
the other hand, represents the contribution from the two-body state
and we call it compositeness $X$:
\begin{equation}
X \equiv \int \frac{d^{3} q}{( 2 \pi )^{3}} 
\langle \psi | \bm{q} \rangle 
\langle \bm{q} | \psi \rangle .
\end{equation}
The elementariness and compositeness satisfy the sum rule
\begin{equation}
1 
= \langle \psi | \psi \rangle 
= Z + X .
\end{equation}
Introducing the momentum space wave function for the two-body state,
$\tilde{\psi}(q)$,
\begin{equation}
\tilde{\psi}( q ) = \langle \bm{q} | \psi \rangle ,
\quad
\tilde{\psi}^{*}( q ) = \langle \psi | \bm{q} \rangle ,
\end{equation}
the compositeness $X$ can be expressed as
\begin{equation}
X = \int \frac{d^{3} q}{( 2 \pi )^{3}} 
\left | \tilde{\psi} ( q ) \right | ^{2} .
\label{eq:Xwavefunction}
\end{equation}
Again, $X$ is real and nonnegative.

For the explicit calculation, we assume the separable form of the
matrix elements of $\hat{V}$ in the momentum space. The matrix
elements are given by
\begin{align}
\langle \bm{q}^{\prime} | \hat{V} | \bm{q} \rangle
= v f^{\ast} ( q^{\prime \, 2} ) f( q^{2} ) , 
\quad 
\langle \bm{q} | \hat{V} | \psi_{0} \rangle
= g_{0} f^{\ast} ( q^{2} ), \quad 
\langle \psi_{0} | \hat{V} | \psi_{0} \rangle
= 0 ,
\end{align}
where $v$ is the interaction strength between the scattering
particles, and $g_{0}$ is the coupling constant of the bare state to
the scattering state.  As we will see later, the one-body state is the
source of the energy dependence of the effective interaction between
the scattering particles.  The matrix element $\langle \psi_{0} |
\hat{V} | \psi_{0} \rangle$ is taken to be zero since it can be
absorbed into $\hat{H}_{0}$ without loss of generality, and throughout
this study the mass of the bare state, $M_{0}$, is not restricted to
be smaller than $M^{\rm th}$ but is allowed to take any value with
this condition.  The form factor $f(q^{2})$ is responsible for the
off-shell momentum dependence of the interaction and suppresses the
high momentum contribution to tame the ultraviolet divergence.  The
normalization is chosen to be $f(0)=1$. The hermiticity of the
Hamiltonian ensures that $v$ is real and
\begin{align}
\langle \psi_{0} | \hat{V} | \bm{q} \rangle
= g^{*}_{0} f( q^{2} ) .
\end{align}
In this study we further assume the time-reversal invariance of the
scattering process, which constraints the interaction, with an
appropriate choice of phases of the states, as
\begin{equation}
\langle \bm{q}^{\prime} | \hat{V} | \bm{q} \rangle
= \langle \bm{q} | \hat{V} | \bm{q}^{\prime} \rangle
= v f ( q^{\prime \, 2} ) f( q^{2} ) , 
\quad 
\langle \bm{q} | \hat{V} | \psi_{0} \rangle
= \langle \psi_{0} | \hat{V} | \bm{q} \rangle
= g_{0} f ( q^{2} ), 
\quad 
\langle \psi_{0} | \hat{V} | \psi_{0} \rangle
= 0 .
\label{eq:potQM} 
\end{equation}
Thus all of the quantities $v$, $g_{0}$, and $f(q^{2})$ are now real.
We emphasize that the assumptions made in the present framework are
just the factorization of the momentum dependence and the
time-reversal invariance of the interaction.  With the
interaction~\eqref{eq:potQM}, we obtain the exact solution of this
system without introducing any further assumptions.

For the separable interaction, the wave function $\tilde{\psi}(q)$ can
be analytically obtained~\cite{Yamaguchi:1954mp}. To this end, we
multiply $\langle \bm{q} |$ and $\langle \psi _{0} |$ to
Eq.~\eqref{eq:Schroedinger}:
\begin{align}
\langle \bm{q} | \hat{H} | \psi \rangle 
&= \left ( M^{\rm th} + \frac{q^{2}}{2 \mu} \right ) 
\tilde{\psi} ( q ) 
+ v f (q^{2}) \int \frac{d^{3} q^{\prime}}{(2 \pi )^{3}} 
f ( q^{\prime \, 2} ) \tilde{\psi} ( q^{\prime} )
+ g_{0} f ( q^{2} ) \langle \psi _{0} | \psi \rangle  
= M_{\rm B} \tilde{\psi} ( q ) , \\
\langle \psi _{0} | \hat{H} | \psi \rangle 
&= M_{0} \langle \psi _{0} | \psi \rangle 
+ g_{0} \int \frac{d^{3} q}{(2 \pi )^{3}} 
f ( q^{2} ) \tilde{\psi} ( q ) 
= M_{\rm B} \langle \psi _{0} | \psi \rangle  ,
\label{eq:Schr_psi0}
\end{align}
where we have inserted Eq.~\eqref{eq:unityQM} between $\hat{V}$ and $|
\psi \rangle$. Eliminating $\langle \psi _{0} | \psi \rangle $ from
these equations, we obtain the \Schr equation for $\tilde{\psi}(q)$ in
an integral form:
\begin{equation}
\left ( M^{\rm th} + \frac{q^{2}}{2 \mu} \right ) 
\tilde{\psi} ( q ) 
+  v^{\rm eff} ( M_{\rm B} ) f(q^{2}) 
\int \frac{d^{3} q^{\prime}}{(2 \pi )^{3}} 
f ( q^{\prime \, 2} ) \tilde{\psi} (q^{\prime})
= M_{\rm B} \tilde{\psi} ( q ) , 
\label{eq:Schr_veff}
\end{equation}
where we have defined the energy-dependent interaction $v^{\rm eff}$
as
\begin{equation}
v^{\rm eff} ( E ) \equiv v 
+ \frac{(g_{0})^{2}}{E - M_{0}} . 
\label{eq:veff}
\end{equation}
Equation~\eqref{eq:Schr_veff} is the single-channel \Schr equation for
the relative motion of the scattering particles under the presence of
the bare state interacting with them by $\hat{V}$.  The effect of the
bare state is incorporated into the energy dependent interaction
$v^{\rm eff}(E)$.

The solution of Eq.~\eqref{eq:Schr_veff} can be obtained as 
\begin{equation}
\tilde{\psi} ( q ) 
= \frac{- c f(q^{2})}{B+q^{2}/(2\mu)} , 
\label{eq:WF_QM-II}
\end{equation}
where we have defined the binding energy $B \equiv M^{\rm th} - M_{\rm
  B} > 0 $ and the normalization constant
\begin{equation}
c \equiv v^{\rm eff} (M_{\rm B})
\int \frac{d^{3} q^{\prime}}{(2 \pi )^{3}} 
f(q^{\prime \, 2}) \tilde{\psi}(q^{\prime}) . 
\label{eq:c}
\end{equation}
In general, Eq.~\eqref{eq:Schr_veff} is an integral equation to
determine the wave function $\tilde{\psi} ( q ) $. For the separable
interaction, however, the integral in Eq.~\eqref{eq:c} (and hence, the
constant $c$) is independent of $q$. In this way, the wave function
$\tilde{\psi} (q)$ is analytically determined by the form factor
$f(q^{2})$ and the constant $c$, which will be determined through the
comparison with the scattering amplitude. Substituting the wave
function~\eqref{eq:WF_QM-II} into Eq.~\eqref{eq:c}, we obtain
\begin{equation}
c = -v^{\rm eff}(M_{\rm B}) \int \frac{d^{3} q}{(2 \pi )^{3}} 
\frac{ [f(q^{2})]^{2}}{B+q^{2}/(2\mu) } c .
\label{eq:cc}
\end{equation}
For the existence of the bound state at $E = M_{\rm B}$,
Eq.~\eqref{eq:cc} should be satisfied with nonzero $c$. The nontrivial
solution can be obtained by
\begin{equation}
1 = v^{\rm eff}(M_{\rm B}) G ( M_{\rm B} ) ,
\label{eq:1-VG_NR}
\end{equation}
where we have introduced a function
\begin{equation}
G (E) = \int \frac{d^{3} q}{(2 \pi )^{3}} 
\frac{[f(q^{2})]^{2}}{E - M^{\rm th} - q^{2}/(2 \mu )} ,
\label{eq:def-GNR}
\end{equation}
which plays an important role in the following discussion and is
called the loop function.  As we will see later, the loop function is
equivalent to the Green function of the free two-body Hamiltonian.  We
note that here and in the following the energy in the denominator of
the loop function is considered to have an infinitesimal positive
imaginary part $i \epsilon$: $E \to E+ i \epsilon$.

The normalization constant $c$ is equal to the square root of the
residue of the scattering amplitude at the pole position of the bound
state.  To prove this, we first represent the compositeness $X$ and
elementariness $Z$ using $c$.  With the explicit form of the wave
function~\eqref{eq:WF_QM-II} and the loop function~\eqref{eq:def-GNR},
the compositeness for the separable interaction can be expressed with
the derivative of the loop function as
\begin{equation}
X = \int \frac{d^{3} q}{(2 \pi )^{3}} 
\left | \tilde{\psi} ( q ) \right | ^{2}
=  - |c|^{2} \left [\frac{d G}{d E}\right ] _{E = M_{\rm B}} .
\label{eq:c_norm}
\end{equation}
We note that both the wave function $\tilde{\psi} (q)$ and the loop
function have the same structure of $1/(E - \hat{H}_{0})$ at $E=M_{\rm
  B}$.  Substituting the wave function into Eq.~\eqref{eq:Schr_psi0},
we obtain
\begin{equation}
\langle \psi_{0}|\psi \rangle
=  \frac{cg_{0}}{M_{\rm B} - M_{0}} 
G ( M_{\rm B} ) , 
\end{equation}
and hence
\begin{equation}
Z = \langle \psi|\psi_{0} \rangle
\langle \psi_{0}|\psi \rangle
=
|c|^{2} G ( M_{\rm B} ) 
\frac{(g_{0})^{2}}{(M_{\rm B} - M_{0})^{2}}
G ( M_{\rm B} )
= -  |c|^{2} \left [
G \frac{d v^{\rm eff}}{d E} G \right ] _{E = M_{\rm B}},
\label{eq:Z_QM}
\end{equation}
where we have used the derivative of Eq.~\eqref{eq:veff}.  We note
that Eqs.~\eqref{eq:c_norm} and \eqref{eq:Z_QM} provide a sum rule
\begin{equation}
1 = - |c|^{2}
\left [ \frac{d G}{d E} 
+ G \frac{d v^{\rm eff}}{d E} G
\right ] _{E = M_{\rm B}} .
\label{eq:norm_QM-II}
\end{equation}

Next the scattering amplitude $t (E)$ is obtained by taking the matrix
element of the $T$-operator for the scattering state $| \bm{q}
\rangle$ with the on-shell condition as $\langle \bm{q}^{\prime} |
\hat{T} | \bm{q} \rangle = t (E) f(q^{\prime \, 2}) f(q^{2}) $ for the
separable interaction.  The $T$-operator satisfies the
Lippmann-Schwinger equation
\begin{equation}
\hat{T} = \hat{V} + \hat{V} \frac{1}{E - \hat{H}_{0}} \hat{T} .
\end{equation}
Inserting the complete set~\eqref{eq:unityQM} between the operators
and eliminating the bare state component from the equation, we obtain
the Lippmann-Schwinger equation for the scattering state as
\begin{equation}
\hat{T} = \hat{V}^{\rm eff}(E) + \hat{V}^{\rm eff}(E) 
\frac{1}{E - \hat{H}_{0}} \hat{T} ,
\label{eq:LS}
\end{equation}
where we have introduced the operator of the effective interaction for
the scattering state as
\begin{equation}
\hat{V}^{\rm eff}(E) \equiv \hat{V} + \hat{V} | \psi_{0} \rangle
\frac{1}{E - M_{0}} \langle \psi_{0}| \hat{V} .
\end{equation}
This operator acts only on the two-body state and its matrix element
leads to $\langle \bm{q}^{\prime} |\hat{V}^{\rm eff}(E)| \bm{q}
\rangle = v^{\rm eff}(E) f(q^{\prime \, 2}) f(q^{2})$.  Taking matrix
element of the two-body state in Eq.~\eqref{eq:LS}, we obtain the
amplitude $t (E)$ algebraically as
\begin{equation}
t(E) = v^{\rm eff}(E) + v^{\rm eff}(E) G (E) t(E) 
= \frac{v^{\rm eff}(E)}{1 - v^{\rm eff}(E) G (E)} ,
\label{eq:LS-Ansatz}
\end{equation}
where $G(E)$ is the same form as Eq.~\eqref{eq:def-GNR}, i.e., the
Green function of the free two-body Hamiltonian.  The bound state
condition~\eqref{eq:1-VG_NR} ensures that the amplitude $t(E)$ has a
pole at $E = M_{\rm B}$.  The residue of the amplitude $t(E)$ at the
pole reflects the properties of the bound state.  The residue turns
out to be real and positive, so we represent the residue as $|g|^{2}$:
\begin{equation}
|g|^{2} 
\equiv \lim _{E \to M_{\rm B}} ( E - M_{\rm B} ) t (E) 
=-\frac{1}{\left [ \frac{d G}{d E} 
+ \frac{1}{(v^{\rm eff})^{2}}\frac{d v^{\rm eff}}{d E} 
\right ] _{E = M_{\rm B}}} .
\label{eq:residue}
\end{equation}
We can interpret $g$ as the coupling constant of the bound state to
the two-body state.  Using the bound state
condition~\eqref{eq:1-VG_NR}, we obtain the relation
\begin{equation}
1 = - |g|^{2}
\left [ \frac{d G}{d E} 
+ G \frac{d v^{\rm eff}}{d E} G
\right ] _{E = M_{\rm B}}. 
\label{eq:dGdE}
\end{equation}
Comparing this with Eq.~\eqref{eq:norm_QM-II}, we find $c = g$ with an
appropriate choice of the phase. 

The equality $c =g$ is also confirmed by the following form of the
$T$-operator:
\begin{equation}
\hat{T}=\hat{V}^{\rm eff}(E) + \hat{V}^{\rm eff}(E) 
\frac{1}{E - \hat{H}_{0} - \hat{V}^{\rm eff}(E)} \hat{V}^{\rm eff}(E) .
\label{eq:amplitudeH}
\end{equation}
As we have seen before, the operator $\hat{H}_{0} + \hat{V}^{\rm eff}$
corresponds to the full Hamiltonian for the two-body system with the
implicit bare state.  Near the bound state pole, the amplitude is
dominated by the pole term in the expansion by the eigenstates of the
full Hamiltonian as
\begin{align}
\lim_{E\to M_{\rm B}} \hat{T} (E)
&\sim 
\hat{V}^{\rm eff} (M_{\rm B})
| \psi\rangle
\frac{1}{E - M_{\rm B}}
\langle \psi | 
\hat{V}^{\rm eff} (M_{\rm B}) ,
\end{align}
and hence, taking the matrix element of the scattering states, we have
\begin{align}
\lim_{E\to M_{\rm B}}t(E)
&\sim 
\int \frac{d^{3} q}{(2 \pi )^{3}} 
\int \frac{d^{3} p}{(2 \pi )^{3}} 
v^{\rm eff} (M_{\rm B})
f(q^{2}) 
\frac{\langle\bm{q} | \psi\rangle\langle \psi| \bm{p}\rangle}
{E-M_{\rm B}}
f(p^{2})
v^{\rm eff} (M_{\rm B}) 
\to \frac{|c|^{2}}{E-M_{\rm B}} ,
\end{align}
where we have used Eq.~\eqref{eq:c}. From the definition of the
residue~\eqref{eq:residue}, this verifies $c =g$.

Here we emphasize that, as seen in Eq.~\eqref{eq:c_norm}, the
compositeness is expressed with the residue of the scattering
amplitude at the pole position and the energy derivative of the loop
function $d G / d E$, and hence the compositeness does not explicitly
depend on the effective interaction $v^{\rm eff}$.\footnote{Since the
  bound state properties are determined by the interaction, the
  compositeness depends implicitly on the effective interaction
  $v^{\rm eff}$.}  Therefore, the compositeness can be obtained solely
with the bound state properties without knowing the details of the
effective interaction, once we fix the loop function, which coincides
with fixing the model space to measure the compositeness via the Green
function of the free two-body Hamiltonian.

We also note that, as seen in Eq.~\eqref{eq:Z_QM}, the elementariness
$Z$ is proportional to the energy derivative of the interaction
$d v^{\rm eff}/d E$ at the bound state energy. This is instructive to
interpret the origin of the elementariness $Z$. In quantum mechanics,
the two-body interaction should not depend on the energy to have an
appropriate normalization.  In the present case, the energy dependence
of $v^{\rm eff}$ stems from the bare state channel $|\psi_{0}\rangle$.
Strong energy dependence of the interaction $v^{\rm eff}$ at the
bound-state pole position emerges when the involved bare state lies
close to the physical bound state, and provides $Z \approx 1$.  This
means that the effect from the bare state is responsible for the
formation of the bound state.  Weak energy dependence, which
corresponds to $Z\approx 0$, can be understood that the bare state
exists far away from the pole position of the physical bound state,
and is insensitive to form the bound state.  In this case, the bound
state is composed dominantly of the scattering channels considered.
This shares viewpoints with Ref.~\cite{Hyodo:2008xr}, where it was
discussed that the energy-dependent Weinberg-Tomozawa term can provide
the effect of the CDD pole~\cite{Castillejo:1955ed}.

\subsection{Coupled scattering channels with multiple bare states}
\label{sec:2-2}

The framework in the last subsection is straightforwardly generalized
to the coupled-channel scattering with multiple one-body bare
states. The eigenstates of the free Hamiltonian $\hat{H}_{0}$ now
include several bare states $| \psi _{a} \rangle$ labeled by $a$ and
two-body scattering states of several channels labeled by $j$.  We
assume that the bound state of which the components we want to examine
is located below the lowest threshold of the two-body channels to make
the state stable.  The normalization and the completeness relation are
given by
\begin{equation}
\langle \bm{q}_{j}^{\prime} | \bm{q}_{k} \rangle 
= (2 \pi )^{3} \delta _{j k} \delta ^{3} (\bm{q}^{\prime} - \bm{q}) , 
\quad
\langle \psi _{a}| \psi _{b} \rangle = \delta _{a b} , 
\quad
\langle \psi _{a}| \bm{q}_{j} \rangle 
= \langle \bm{q}_{j} | \psi _{a} \rangle =0 , 
\label{eq:unityQMcoupledstates}
\end{equation}
\begin{equation}
1 
= \sum _{a} | \psi _{a} \rangle \langle \psi _{a} | 
+ \sum _{j} \int \frac{d^{3} q}{(2 \pi )^{3}} 
| \bm{q}_{j} \rangle  \langle \bm{q}_{j} | .
\label{eq:unityQMcoupled}
\end{equation}
The matrix elements of the interaction are 
\begin{equation}
  \langle \bm{q}_{j}^{\prime} | \hat{V} | \bm{q}_{k} \rangle
  = v_{j k} f_{j} ( q^{\prime \, 2} ) f_{k} ( q^{2} ) , 
  \quad 
  \langle \bm{q}_{j} | \hat{V} | \psi _{a} \rangle 
  = \langle \psi _{a} | \hat{V} | \bm{q}_{j} \rangle 
  = g_{0}^{a, j} f_{j} ( q^{2} ) , 
  \quad 
  \langle \psi_{a} | \hat{V} | \psi_{b} \rangle
  = 0 ,
\end{equation}
where, due to the time-reversal invariance, $v_{j k}$ is a real
symmetric matrix and $g_{0}^{a, j}$ and $f_{j}(q^{2})$ are real with
an appropriate choice of phases of states.  The total normalization of
the bound state wave function now leads to
\begin{equation}
 1 = \sum _{a} Z_{a} +\sum_{j}X_{j} ,
 \label{eq:normalizationcoupled}
\end{equation}
with the elementariness
\begin{equation}
Z_{a} \equiv \langle \psi | \psi _{a} \rangle 
\langle \psi _{a} | \psi \rangle ,
\end{equation}
and the compositeness given by the wave function for each channel
\begin{equation}
X_{j} \equiv \int \frac{d^{3} q}{( 2 \pi )^{3}} 
\left | \tilde{\psi}_{j} ( q ) \right | ^{2} , 
\label{eq:Xj}
\end{equation}
where
\begin{equation}
\tilde{\psi}_{j} ( q ) = \langle \bm{q}_{j} | \psi \rangle ,
\quad
\tilde{\psi}_{j}^{*} ( q ) = \langle \psi | \bm{q}_{j} \rangle .
\end{equation}

We follow the same procedure as the single channel case; incorporating
the one-body bare states to the effective interaction for the two-body
states, we obtain the coupled \Schr equation as
\begin{equation}
\left ( M_{j}^{\rm th} + \frac{q^{2}}{2 \mu _{j}} \right ) 
\tilde{\psi}_{j} ( q ) 
+ \sum _{k} v_{j k}^{\rm eff} ( M_{\rm B} ) f_{j} (q^{2}) 
\int \frac{d^{3} q^{\prime}}{(2 \pi )^{3}} 
f_{k} ( q^{\prime \, 2} ) \tilde{\psi}_{k} (q^{\prime})
= M_{\rm B} \tilde{\psi}_{j} ( q ) , 
\end{equation}
where $M_{j}^{\rm th}$ and $\mu _{j}$ are the threshold and the
reduced mass in channel $j$, respectively, and we have defined the
energy-dependent effective interaction as
\begin{equation}
v_{j k}^{\rm eff} ( E ) \equiv v_{j k} 
+ \sum _{a} \frac{g_{0}^{a,j} g_{0}^{a,k}}{E - M_{a}} ,
\label{eq:veff_jk}
\end{equation}
which is a real symmetric matrix for a real energy and $M_{a}$ is the
mass of the bare state. The \Schr equation can be solved algebraically
again for the separable interaction:
\begin{equation}
\tilde{\psi}_{j} ( q ) 
= \frac{- c_{j} f_{j}(q^{2})}{ B_{j} + q^{2}/(2\mu_{j}) } ,  
\label{eq:psi_j}
\end{equation}
where $B_{j} \equiv M_{j}^{\rm th} - M_{\rm B}$ is the binding energy
measured from the $j$-channel threshold.  The normalization constant
is given by
\begin{equation}
c_{j} \equiv \sum _{k} v_{j k}^{\rm eff} (M_{\rm B})
\int \frac{d^{3} q}{(2 \pi )^{3}} 
f_{k}(q^{2}) \tilde{\psi}_{k}(q) .
\label{eq:cj}
\end{equation}
With substitution of Eq.~\eqref{eq:psi_j} to Eq.~\eqref{eq:cj}, the
bound state condition for nonzero $c_{j}$ can be summarized as
\begin{equation}
\det \left [ 1 - v^{\rm eff} ( M_{\rm B} ) G ( M_{\rm B} ) \right ] = 0 ,
\label{eq:1-VG_NR-II}
\end{equation}
with the loop function
\begin{equation}
G_{j} ( E ) =
\int \frac{d^{3} q}{(2 \pi )^{3}} 
\frac{[f_{j} ( q^{2} )]^{2}}
{E - M_{j}^{\rm th} - q^{2}/(2 \mu _{j})} ,
\end{equation}
which is diagonal with respect to the channel index.

The coupled-channel scattering equation is in the matrix form
\begin{equation}
t ( E ) = \left [ 1 - v^{\rm eff} (E) G (E) \right ] ^{-1} v^{\rm eff} (E) ,
\end{equation}
where the channel index runs through only the scattering channels,
since the one-body bare states are incorporated into the effective
interaction $v^{\rm eff}$.  Equation~\eqref{eq:1-VG_NR-II} ensures the
existence of the bound state pole at $E = M_{\rm B}$.  The residue of the
amplitude at the pole, which is real for the bound state, is
interpreted as the product of the coupling constants\footnote{Since an
  interaction of a symmetric matrix $v_{j k}^{\rm eff}$ leads to a
  symmetric $t$-matrix, $t_{j k} = t_{k j}$, the residue of the
  $t$-matrix is also symmetric and can be factorized as $g_{j}
  g_{k}$.}
\begin{equation}
g_{j} g_{k} = \lim _{E \to M_{\rm B}} ( E - M_{\rm B} ) t_{j k} (E) .
\end{equation}
On the other hand, using the coupled-channel version of
Eq.~\eqref{eq:amplitudeH}, the amplitude near the bound state pole is
given by
\begin{align}
\lim_{E\to M_{\rm B}}t_{j k}(E)
&\sim 
\sum_{l,m}
\int \frac{d^{3} q}{(2 \pi )^{3}} 
\int \frac{d^{3} p}{(2 \pi )^{3}} 
v_{jl}^{\rm eff} (M_{\rm B})
f_{l}(q^{2}) 
\frac{\langle\bm{q}_{l} | \psi\rangle\langle \psi| \bm{p}_{m}\rangle}
{E-M_{\rm B}}
f_{m}(p^{2})
v_{mk}^{\rm eff} (M_{\rm B}) 
\nonumber \\ &
\to \frac{c_{j}c_{k}^{*}}{E-M_{\rm B}} ,
\end{align}
which shows that $c_{j}=g_{j}$ with an appropriate choice of the
phase.

Now the compositeness in channel $j$ can be expressed as
\begin{equation}
X_{j} = \int \frac{d^{3} q}{(2 \pi )^{3}} 
\left | \tilde{\psi}_{j} ( q ) \right | ^{2}
=  - |c_{j}|^{2} \left [\frac{d G_{j}}{d E}\right ] _{E = M_{\rm B}} 
=  - |g_{j}|^{2} \left [\frac{d G_{j}}{d E}\right ] _{E = M_{\rm B}} .
\label{eq:c_normcoupled}
\end{equation}
The overlap of the bound state wave function with the bare state $\psi
_{a}$ is given by
\begin{equation}
\langle \psi_{a}|\psi \rangle
= \frac{1}{M_{\rm B} - M_{a}} 
\sum_{j} c_{j} g_{0}^{a,j} G_{j} ( M_{\rm B} ) , 
\end{equation}
and thus we obtain 
\begin{equation}
Z_{a} = \langle \psi | \psi _{a} \rangle 
\langle \psi _{a} | \psi \rangle 
= \sum_{j,k}
c_{k}c_{j}^{*}
 G_{j} ( M_{\rm B} ) 
G_{k} ( M_{\rm B} )
\frac{g_{0}^{a,j} g_{0}^{a,k}}{(M_{\rm B} - M_{a})^{2}} . 
\end{equation}
The total elementariness $Z \equiv \sum _{a} Z_{a}$, which contains
all contributions from the implicit channels, is
\begin{equation}
Z \equiv \sum _{a} Z_{a} =
\sum_{j,k}
c_{k}c_{j}^{*}
 G_{j} ( M_{\rm B} ) 
G_{k} ( M_{\rm B} )
\sum _{a} \frac{g_{0}^{a,j} g_{0}^{a,k}}{(M_{\rm B} - M_{a})^{2}}
= - \sum_{j,k}
g_{k}g_{j}
 \left [
G_{j} \frac{d v^{\rm eff}_{j k}}{d E} G_{k} \right ] _{E = M_{\rm B}} .
\label{eq:Z_QM-II}
\end{equation}
From the normalization~\eqref{eq:normalizationcoupled}, we obtain the sum 
rule
\begin{equation}
- \sum _{j,k} g_{k}  g_{j}
\left [ \delta _{j k} \frac{d G_{j}}{d E} 
+ G_{j} \frac{d v_{j k}^{\rm eff}}{d E} G_{k} 
\right ] _{E = M_{\rm B}} = 1 . 
\label{eq:norm_QM-III}
\end{equation}
This corresponds to the nonrelativistic counterpart of the generalized
Ward identity derived in Ref.~\cite{Sekihara:2010uz}.  We note that
the sum rule~\eqref{eq:norm_QM-III} as the normalization of the wave
function can be obtained by the explicit treatment of both the
two-body states and the one-body bare states, which complement the
discussion of the bound-state wave function with an energy-independent
separable interaction done in Ref.~\cite{Gamermann:2009uq}.

So far, we have regarded the components coming from the one-body bare
states as the elementariness.  On the other hand, sometimes it happens
that some of the two-body channel thresholds are so high enough that
these channels may play a minor role.  In such a case, these channels
can be also included into implicit channels of the effective
interaction $v^{\rm eff}$ by, {\it e.g.}, the Feshbach
method~\cite{Feshbach:1958nx, Feshbach:1962ut}.  These implicit
channels also provide energy dependence of the effective interaction
which acts on the reduced model space (see also
Ref.~\cite{Aceti:2014ala}), and accordingly we are allowed to
interpret the contributions coming from these channels as the
elementariness.  For instance, the implementation of a scattering
channel $N$ into the effective interaction can be done by replacing
$v^{\rm eff}$ as:
\begin{equation}
w_{j k} ( E )
= v_{j k}^{\rm eff} 
+ v_{j N}^{\rm eff} \frac{G_{N} ( E )}{1 - v_{N N}^{\rm eff} G_{N} ( E )}
v_{N k}^{\rm eff} , 
\quad 
j, \, k \ne N ,
\end{equation}
where the $N$-th channel has been absorbed in the effective
interaction $w_{j k}$ in the same manner as in~\cite{Hyodo:2007jq}.
In this case the elementariness $Z^{w}$ may be able to be calculated
by the derivative of the effective interaction $w_{j k}$ as
\begin{equation}
Z^{w} = - \sum _{j , k \ne N} 
g_{k} g_{j} \left [ G_{j} \frac{d w_{j k}}{d E} 
G_{k} \right ] _{E = M_{\rm B}} .
\label{eq:Zw}
\end{equation}
Interestingly, the elementariness $Z^{w}$ can be expressed as the sum
of the elementariness with the full two-body channels, $Z$, and the
$N$-th channel compositeness $X_{N}$, namely,
\begin{equation}
Z^{w} = Z + X_{N} , 
\label{eq:ZnoN}
\end{equation}
with
\begin{equation}
Z = - \sum _{j , k} 
g_{k} g_{j} \left [ G_{j} \frac{d v^{\rm eff}_{j k}}{d E} 
G_{k} \right ] _{E = M_{\rm B}} ,
\quad 
X_{N} = - g_{N}^{2} 
\left [\frac{d G_{N}}{d E}\right ] _{E = M_{\rm B}} .
\end{equation}
The proof is shown in Appendix~\ref{sec:A}.  In this way, the
elementariness can be redefined by Eq.~\eqref{eq:Zw}.  With this
expression the elementariness measures contributions coming from both
one-body bare states and two-body channels which are implemented into
the effective interaction and do not appear as explicit degrees of
freedom.

At the end of this subsection, we mention that our formulation of the
compositeness and elementariness can be applied to any separable
interactions with arbitrary energy dependence by interpreting that the
energy dependence on the effective interaction comes from the implicit
channels.  Actually, when the compositeness and elementariness are
formulated with multiple one-body bare states, all of these bare
states are included in the effective two-body interaction $v^{\rm
  eff}(E)$ and the total elementariness is calculated as the sum of
each bare-state contribution, which is essentially the derivative of
the effective two-body interaction as in Eq.~\eqref{eq:Z_QM-II}.  It
is important that in this case we can produce any energy dependent
interactions with suitable bare states.  In order to see this, for
instance, we assume that the mass of a bare state is large enough, and
by expanding the bare-state term in the effective interaction as
\begin{equation}
  \frac{1}{E - M_{0}} 
  = - \frac{1}{M_{0}} \left ( 1 + \frac{E}{M_{0}} 
    + \cdots \right ) , 
\end{equation}
we have polynomial energy dependence in the effective interaction.
This fact enables us to apply the formulae of the compositeness and
elementariness to interactions with an arbitrary energy dependence.
This is the foundation of the analysis of physical hadronic resonances
in Sec.~\ref{sec:3}.

\subsection{Weak binding limit and threshold parameters}
\label{sec:2-3}

In this subsection, we consider the weak binding limit to derive the
Weinberg's compositeness condition~\cite{Weinberg:1965zz} on the
scattering length $a$ and the effective range $r_{e}$. This ensures
that the expression for the compositeness in this paper correctly
reproduces the model-independent result of Ref.~\cite{Weinberg:1965zz}
in the weak binding limit.  For simplicity we consider a system with one
scattering channel like in Sec.~\ref{sec:2-1}.

In the single-channel problem, the elastic scattering amplitude
$\mathcal{F} (E)$ is written with the $t$-matrix $t(E)$ given in
Eq.~\eqref{eq:LS-Ansatz} as
\begin{equation}
\mathcal{F} (E)
= -\frac{1}{(2\pi)^{3}}(2\pi)^{2}\mu t(E)[f(k^{2})]^{2} ,
\end{equation}
with $k \equiv \sqrt{2 \mu (E - M^{\rm th})}$.  The scattering length
$a$ is defined as the value of the scattering amplitude at the
threshold:
\begin{equation}
a \equiv - \mathcal{F} (M^{\rm th}) =\frac{\mu}{2 \pi} t ( M^{\rm th} ) 
= \frac{\mu}{2 \pi} \frac{1}{v^{-1} ( M^{\rm th} ) - G ( M^{\rm th} )} ,
\end{equation}
where we have abbreviated $v^{\rm eff}$ as $v$ for simplicity. Now we
perform the expansion in terms of the energy $E$ around $M_{\rm B}$ by
considering $B=M^{\rm th}-M_{\rm B}$ to be small. To expand the
denominator, we write
\begin{equation}
v^{-1} ( M^{\rm th} ) 
= v^{-1} ( M_{\rm B} ) 
+ B 
\left [
\frac{d v^{-1}}{d E} 
\right ] _{E = M_{\rm B}}
+\Delta v^{-1} ,
\label{eq:v_exp} 
\end{equation}
\begin{equation}
G ( M^{\rm th} ) 
= G ( M_{\rm B} ) 
+ B \left [\frac{d G}{d E} \right ] _{E = M_{\rm B}}
+ \Delta G , 
\label{eq:G_exp}
\end{equation}
where we have defined 
\begin{equation}
\Delta v^{-1} \equiv \sum _{n=2}^{\infty} \frac{B^{n}}{n !} 
\left [\frac{d^{n} v^{-1}}{d E^{n}}\right ] _{E = M_{\rm B}} ,
\quad 
\Delta G \equiv \sum _{n=2}^{\infty} \frac{B^{n}}{n !} 
\left [\frac{d^{n} G}{d E^{n}}\right ] _{E = M_{\rm B}} .
\end{equation}
Here we allow arbitrary energy dependence for $v$, as stated in the
end of the last subsection, but assume that the effective range
expansion is valid up to the energy of the bound state, which is a
precondition for the formula in Ref.~\cite{Weinberg:1965zz}.  In this
case there should exist no singularity of $v^{-1}(E)$ between
$E=M_{\rm B}$ and $M^{\rm th}$ and expansion~\eqref{eq:v_exp} is
safely performed up to the threshold, and hence $\Delta
v^{-1}=\mathcal{O}(B^{2})$.  Otherwise the singularity of $v^{-1}(E)$
around the threshold spoils the effective range expansion, as the
divergence of $v^{-1}$ leads to the existence of the CDD pole.  As a
result, with the bound state condition~\eqref{eq:1-VG_NR}, the
scattering length is now given by
\begin{equation}
a =
\frac{\mu}{2 \pi} 
\left(
B 
\left [
\frac{d v^{-1}}{d E} 
-\frac{d G}{d E}
\right ] _{E = M_{\rm B}}
-\Delta G
+ \mathcal{O}(B^{2}) %\Delta v^{-1}
\right)^{-1} .
\label{eq:aexact}
\end{equation}

The first term in the parenthesis in Eq.~\eqref{eq:aexact} is
calculated as
\begin{align}
B 
\left [
\frac{d v^{-1}}{d E} 
-\frac{d G}{d E}
\right ] _{E = M_{\rm B}}
&=-B 
\left [
G^{2}
\frac{d v}{d E} 
+\frac{d G}{d E}
\right ] _{E = M_{\rm B}} \nonumber \\
&=\frac{B}{|g|^{2}} \nonumber \\
&=-\frac{B}{X}
\left [
\frac{d G}{d E}
\right ] _{E = M_{\rm B}} \nonumber \\
&=\frac{B}{X}
\int \frac{d^{3} q}{(2 \pi )^{3}} 
\frac{[f(0)]^{2}+\mathcal{O}(q^{2})}{[B + q^{2}/ (2 \mu )]^{2}} 
\nonumber\\
&=\frac{\mu}{4\pi X} \frac{1}{R} 
+\mathcal{O}(B) ,
\label{eq:firstterm}
\end{align}
where we have used Eqs.~\eqref{eq:1-VG_NR}, \eqref{eq:dGdE},
\eqref{eq:c_norm}, and the normalization $f(0)=1$, and we have defined
$R \equiv 1 / \sqrt{2 \mu B}$ in the last line. To evaluate $\Delta
G$, we first note that
\begin{align}
\left [ \frac{d^{n} G}{d E^{n}} \right ] _{E = M_{\rm B}}
&=  \int \frac{d^{3} q}{(2 \pi )^{3}} 
\frac{( - 1 )^{n} n ! [f(q^{2})]^{2}}
{[M_{\rm B} - M^{\rm th} - q^{2}/ (2 \mu )]^{n+1}}
\nonumber \\
&= - n ! \int \frac{d^{3} q}{(2 \pi )^{3}} 
\frac{[f(q^{2})]^{2}}{[B + q^{2}/ (2 \mu )]^{n+1}} 
\nonumber \\
&=  - \frac{n !}{B^{n}} \frac{2 \mu}{R}
\int \frac{d^{3} q^{\prime}}{(2 \pi )^{3}} 
\frac{[f(2 \mu B q^{\prime \, 2})]^{2}}{(q^{\prime \, 2} + 1 )^{n+1}} , 
\end{align}
where $\bm{q}^{\prime} \equiv R \bm{q}$. Thus summing up all
contributions we have
\begin{align}
\Delta G
&=  -\sum _{n=2}^{\infty} \frac{2 \mu}{R}
\int \frac{d^{3} q^{\prime}}{(2 \pi )^{3}} 
\frac{[f(0)]^{2}}{(q^{\prime \, 2} + 1 )^{n+1}} 
+\mathcal{O}(B) 
\nonumber \\
&= - \frac{\mu}{\pi ^{2} R}  
\int _{0}^{\infty} d x \, x^{2} \sum _{n=2}^{\infty} 
\frac{1}{( x^{2} + 1 )^{n+1}} + \mathcal{O}(B) 
\nonumber \\
&= -  \frac{\mu}{4 \pi} \frac{1}{R} +\mathcal{O}(B) ,
\label{eq:DeltaG}
\end{align}
where we have used the summation relation
\begin{equation}
\sum _{n=2}^{\infty} \frac{1}{( x^{2} + 1 )^{n+1}} 
= \frac{1}{x^{2} ( x^{2} + 1 )^{2}} 
\quad ( x \ne 0 ) .
\end{equation}

As a consequence, we obtain the expression of the scattering length in
terms of the compositeness $X$ from Eqs.~\eqref{eq:aexact},
\eqref{eq:firstterm} and \eqref{eq:DeltaG}:
\begin{equation}
a = \frac{\mu}{2 \pi} \left ( 
\frac{\mu}{4 \pi X} \frac{1}{R}
+ \frac{\mu}{4 \pi} \frac{1}{R} 
+\mathcal{O}(B)
\right ) ^{-1}
= R 
\frac{2 X}{1 + X} +\mathcal{O}(B^{0}) ,
\label{eq:weakbinding}
\end{equation}
which agrees with the result in Ref.~\cite{Weinberg:1965zz} with $X =
1 - Z$. It is important that in the weak binding limit the details of
the form factor $f(q^{2})$ are irrelevant to the determination of the
compositeness of the bound state from the scattering length of two
constituents.  In contrast, the correction terms of $\mathcal{O}
(B^{0})$ depend on the explicit form of the function $f(q^{2})$.

Because we have assumed that the bound state pole lies within the
valid region of the effective range approximation, the relation
between the scattering length and the effective range is given
by\footnote{The relation~\eqref{eq:a_re} can be obtained from the
  condition $\mathcal{F}^{-1} ( k ) = - 1/a - i k + r_{e} k^{2} / 2 =
  0$ at $k = i / R$.}
\begin{equation}
r_{e} = 2 R
\left ( 1 - \frac{R}{a} \right ) 
\label{eq:a_re}
\end{equation}
Comparing it with Eq.~\eqref{eq:weakbinding}, we find
\begin{equation}
r_{e} = R 
\frac{X-1}{X}
+ \mathcal{O}(B^{0}) .
\end{equation}
This again corresponds to the expression in
Ref.~\cite{Weinberg:1965zz}.

In this way, the structure of the bound state can be determined from
$a$ and $r_{e}$ unambiguously in the weak binding limit.  This means
that, in principle, tuning $a$ and $r_{e}$ could lead to arbitrary
structure of the bound state.  It is however shown in
Ref.~\cite{Hanhart:2014ssa} that the bound state with $Z \sim 0$
naturally appears when the state exists near the threshold, and a
significant fine tuning is required to realize $Z \sim 1$ in this
small binding region.  This behavior can be understood by considering
the value of $Z$ in the exact $B \to 0$ limit.  Actually, the value of
$Z$ is shown to vanish in the $B \to 0$ limit, as far as the bound
state pole exist in the scattering amplitude~\cite{Hyodo:2014bda}.  It
is therefore natural to expect that the bound state should be $Z \sim
0$ in the small binding region.

\subsection{Generalization to resonances}
\label{sec:2-4}

Now we generalize our argument to a resonance state. We first
introduce the Gamow state~\cite{Gamow:1928zz} denoted as $| \psi )$ to
express the resonance state.  The eigenvalue of the Hamiltonian is
allowed to be complex for the Gamow state:
\begin{equation}
\hat{H} | \psi ) 
= \left ( M_{\rm R} - i \frac{\Gamma _{\rm R}}{2} \right ) | \psi ) .
\end{equation}
Here $M_{\rm R}$ and $\Gamma _{\rm R}$ are the mass and width of the
resonance state, respectively. The state with a complex eigenvalue cannot 
be normalized in the ordinary sense. To establish the normalization, we 
define the corresponding bra-state as the complex conjugate of the Dirac 
bra-state:
\begin{equation}
( \psi | \equiv \langle \psi ^{\ast}| ,
\label{eq:resonancebra}
\end{equation}
which was firstly introduced to describe unstable
nuclei~\cite{Hokkyo:1965zz, Berggren:1968zz,Romo:1968zz}.  As a
consequence, the eigenvalue of the Hamiltonian is the same with the
ket vector:\footnote{The eigenvectors $|\psi^{*})$ and
  $(\psi^{*}|=\langle \psi |$ have the eigenvalue $M_{\rm
    R}+i\Gamma_{\rm R}/2$.}
\begin{equation}
( \psi | \hat{H} 
= \left ( M_{\rm R} - i \frac{\Gamma _{\rm R}}{2} \right ) ( \psi | .
\label{eq:Schr_res_bra}
\end{equation}
These eigenvectors can be normalized as
\begin{equation}
( \psi | \psi ) = 1 .
\label{eq:RWFnorm}
\end{equation}
With the same eigenstates of the free Hamiltonian in
Eqs.~\eqref{eq:unityQMcoupledstates} and \eqref{eq:unityQMcoupled}, we
can decompose this normalization as
\begin{equation}
1
= \sum _{a} ( \psi | \psi _{a} \rangle \langle \psi _{a} | \psi ) 
+ \sum _{j} \int \frac{d^{3} q}{(2 \pi )^{3}}
( \psi | \bm{q}_{j} \rangle \langle \bm{q}_{j} | \psi ) 
= \sum _{a} Z_{a} + \sum _{j} X_{j} ,
\label{eq:normalizationcoupledres}
\end{equation}
where we have defined the elementariness $Z_{a}$ and compositeness
$X_{j}$ as
\begin{equation}
Z_{a} \equiv ( \psi | \psi _{a} \rangle \langle \psi _{a} | \psi ) ,
\quad 
X_{j} \equiv \int \frac{d^{3} q}{(2 \pi )^{3}}
( \psi | \bm{q}_{j} \rangle \langle \bm{q}_{j} | \psi ) .
\end{equation}
In addition, we define the momentum space wave function
$\tilde{\psi}_{j} ( q ) \equiv \langle \bm{q}_{j} | \psi )$. It
follows from Eqs.~\eqref{eq:resonancebra} and \eqref{eq:Schr_res_bra}
that
\begin{equation}
( \psi | \bm{q}_{j} \rangle = \langle \bm{q}_{j} | \psi )
=\tilde{\psi}_{j} ( q ) .
\label{eq:res_wf}
\end{equation}
The compositeness is then given by
\begin{equation}
X_{j} = \int \frac{d^{3} q}{(2 \pi )^{3}} 
\left [ \tilde{\psi}_{j} ( q ) \right ] ^{2} .
\end{equation}
In contrast to Eq.~\eqref{eq:Xj}, where $X_{j}$ is given by the
absolute value squared, the compositeness of the resonance is given by
the complex number squared.  This is also the case for $Z_{a}$,
because $( \psi | \psi _{a} \rangle = \langle \psi _{a} | \psi ) \neq
\langle \psi _{a} | \psi )^{*}$. In this way, $Z_{a}$ and $X_{j}$ are
in general complex, and the probabilistic interpretation of $Z_{a}$
and $X_{j}$ is not guaranteed.

To determine the wave function, we solve the \Schr equation
\begin{align}
&\quad \left(M_{\rm R} - i \frac{\Gamma _{\rm R}}{2}\right) 
\tilde{\psi}_{j} ( q ) \nonumber \\
&= \left ( M^{\rm th}_{j} + \frac{q^{2}}{2 \mu_{j}} \right ) 
\tilde{\psi}_{j} ( q ) 
+ \sum_{k} v_{j k} f_{j} (q^{2}) \int \frac{d^{3} q^{\prime}}{(2 \pi )^{3}} 
f_{k} ( q^{\prime \, 2} ) \tilde{\psi}_{k} ( q^{\prime} )
+ \sum _{a} g_{0}^{a,j} f_{j} ( q^{2} ) \langle \psi _{a} | \psi )  , 
\end{align}
and
\begin{align}
\left(M_{\rm R} - i \frac{\Gamma _{\rm R}}{2}\right) 
 \langle \psi _{a} | \psi ) 
&= M_{a} \langle \psi _{a} | \psi ) 
+ \sum_{k} g_{0}^{a, k} \int \frac{d^{3} q}{(2 \pi )^{3}} 
f_{k} ( q^{2} ) \tilde{\psi}_{k} ( q )  .
\label{eq:psi0_res}
\end{align}
Eliminating $\langle \psi _{a} | \psi ) $, we obtain
\begin{equation}
\tilde{\psi}_{j} ( q ) 
= \frac{-  c_{j} f_{j}(q^{2})}{ M_{j}^{\rm th} - M_{\rm R} 
+ i \Gamma _{\rm R}/2 + q^{2}/(2\mu_{j}) } ,  
\end{equation}
with the normalization constant
\begin{equation}
c_{j} \equiv \sum _{k} v_{j k}^{\rm eff} 
(M_{\rm R} - i \Gamma _{\rm R}/2)
\int \frac{d^{3} q}{(2 \pi )^{3}} 
f_{k}(q^{2}) \tilde{\psi}_{k}(q)  ,
\end{equation}
where $ v_{j k}^{\rm eff}(E)$ is defined in the same way with
Eq.~\eqref{eq:veff_jk}.  The condition for nonzero $c_{j}$ is
\begin{equation}
\det [ 1 - v^{\rm eff} ( M_{\rm R} - i \Gamma _{\rm R}/2 ) 
G ( M_{\rm R} - i \Gamma _{\rm R}/2 ) ] = 0 .
\label{eq:resonancecond}
\end{equation}
This is the condition for the resonance pole at $E=M_{\rm R} - i
\Gamma _{\rm R}/2$. We note that the loop function in the complex
energy plane is defined on the $2^{n}$-sheeted Riemann surface for an
$n$-channel problem. The resonance pole can exist in any sheets,
except for the one which is reached by choosing the first sheet for
all channels. The most relevant Riemann sheet for the scattering
amplitude at a given energy is reached by choosing the first sheet for
the closed channels and the second sheet for the open channels. In the
following, we concentrate on the poles in this Riemann sheet, while
the framework is in principle applicable to the complex poles in the
other Riemann sheets.

Also for the resonance pole, the residue of the scattering amplitude
is interpreted as product of the coupling constants $g_{j} g_{k}$:
\begin{equation}
g_{j} g_{k} = \lim _{E \to M_{\rm R} - i \Gamma _{\rm R}/2} 
( E - M_{\rm R} + i \Gamma _{\rm R}/2 ) t_{j k} (E) ,
\end{equation}
where the complex conjugate should not be taken for the coupling
constant $g_{k}$ since $t_{j k}$ is symmetric: $t_{j k} = t_{k j}$.
In contrast to the bound states, the coupling constant $g_{j}$ is in
general complex.  The amplitude near the resonance pole is also given
by
\begin{align}
\lim_{E \to M_{\rm R} - i \Gamma _{\rm R}/2}t_{j k}(E)
&\sim 
\sum_{l,m}
\int \frac{d^{3} q}{(2 \pi )^{3}} 
\int \frac{d^{3} p}{(2 \pi )^{3}} 
v_{j l}^{\rm eff} (E)
f_{l}(q^{2}) 
\frac{\langle\bm{q}_{l} | \psi )( \psi| \bm{p}_{m}\rangle}
{E - M_{\rm R} + i \Gamma _{\rm R}/2}
f_{m}(p^{2})
v_{mk}^{\rm eff} (E) \nonumber \\
&\to \frac{c_{j}c_{k}}{ E - M_{\rm R} + i \Gamma _{\rm R}/2 } ,
\label{eq:resonancecoupling}
\end{align}
thus we find $c_{j}=g_{j}$. The compositeness in channel $j$ is then 
given by
\begin{equation}
X_{j} = \int \frac{d^{3} q}{(2 \pi )^{3}} 
\left [ \tilde{\psi}_{j} ( q ) \right ] ^{2} 
= - g_{j}^{2} 
\left [\frac{d G_{j}}{d E} 
\right ] _{E = M_{\rm R} - i \Gamma _{\rm R} / 2} .
\label{eq:Rcomp}
\end{equation}
The loop function in the complex energy plane should be evaluated by 
choosing the Riemann sheets consistently with the choice to obtain the 
pole condition~\eqref{eq:resonancecond}.

From Eq.~\eqref{eq:psi0_res} and its counterpart coming from
  Eq.~\eqref{eq:Schr_res_bra}, we obtain
\begin{align}
\langle \psi_{a}|\psi )
= ( \psi|\psi_{a} \rangle
=  \sum_{j}\frac{c_{j} g_{0}^{a,j}}{M_{\rm R} - i \Gamma _{\rm R} / 2 - M_{a}} 
G_{j} ( M_{\rm R} - i \Gamma _{\rm R} / 2 ) , 
\end{align}
so the total elementariness $Z \equiv \sum _{a} Z_{a}$ is obtained as
\begin{equation}
Z \equiv \sum _{a} Z_{a} = - \sum _{j, k} g_{k}g_{j}  \left [ G_{j} 
\frac{d v_{j k}^{\rm eff}}{d E} G_{k} 
\right ] _{E = M_{\rm R} - i \Gamma _{\rm R} / 2}.
\end{equation}
Using Eqs.~\eqref{eq:normalizationcoupledres}, we obtain
\begin{equation}
- \sum _{j, k} g_{k} g_{j}  \left [ 
\delta _{j k} \frac{d G_{j}}{d E}
+G_{j} \frac{d v_{j k}^{\rm eff}}{d E} G_{k} 
\right ] _{E = M_{\rm R} - i \Gamma _{\rm R} / 2} = 1 .
\label{eq:res_sum-rule}
\end{equation}
This corresponds to the nonrelativistic counterpart of the generalized
Ward identity for resonance states derived in
Ref.~\cite{Sekihara:2010uz}.  The special case of $Z=0$ of
Eq.~\eqref{eq:res_sum-rule} is obtained in
Ref.~\cite{YamagataSekihara:2010pj} by using an energy-independent
separable interaction without the bare-state contribution.  Here we
mention that we should obtain the same results in appropriate ways to
treat resonance states such as the complex scaling
method~\cite{Aoyama:2006CSM}.\footnote{In Sec.~\ref{sec:3-2} we will
  compare the structure of $\Lambda (1405)$ in the present framework
  with that in the complex scaling method.}

By definition, the compositeness for the resonance state becomes
complex. Therefore, strictly speaking, it cannot be interpreted as a
probability of finding the two-body component. Nevertheless, because
it represents the contribution of the channel wave function to the
total normalization, the compositeness $X_{j}$ will have an important
piece of information on the structure of the resonance.  For instance,
consider a resonance such that the real part of a single $X_{j}$ is
close to unity with small imaginary part, and all the other components
have small absolute values. In this case, the resonance wave function
is considered to be similar to that of the bound state dominated by
the $j$-th channel. It is therefore natural to interpret the resonance
state in this case is dominated by the component of the channel
$j$. In general, however, all $X_{j}$ and $Z$ can be arbitrary complex
numbers constrained by Eq.~\eqref{eq:normalizationcoupledres}. The
interpretation of the structure of such a state from $X_{j}$ and $Z$
is not straightforward.

\subsection{Relativistic covariant formulation}
\label{sec:2-5}

Finally we consider the coupled-channel two-body scattering in a
relativistic form. Here we do not consider the intermediate states
with more than two particles but simply solve the two-body wave
equation.\footnote{In general relativistic field theory, there are
  infinitely many diagrams which contribute to the scattering
  amplitude. The present formulation picks up the summation of the
  $s$-channel two-body loop diagrams, which is the most dominant
  contribution in the nonrelativistic limit.}  To describe the wave
function of the resonances, we extract the relative motion of the
two-body system from a relativistic scattering equation with a
three-dimensional reduction~\cite{Mandelzweig:1986hk, Wallace:1989nm}.

According to Appendix~\ref{sec:B}, we introduce the state $|
\bm{q}_{j}^{\text{co}} \rangle$ as the two-body scattering state of
the particles with masses $m_{j}$ and $M_{j}$ and the relative
momentum $\bm{q}$, and its normalization is fixed as
\begin{equation}
\langle \bm{q}_{j}^{\prime \, \text{co}} | \bm{q}_{k}^{\text{co}}  \rangle
= \frac{2 \omega _{j} (\bm{q}) \Omega _{j} (\bm{q})}
{\sqrt{s_{q j}}} (2 \pi )^{3} \delta _{j k} 
\delta ^{3} (\bm{q}^{\prime} - \bm{q}) , 
\label{eq:qco_norm}
\end{equation}
where $s_{q j} \equiv [\omega _{j} (\bm{q}) + \Omega _{j}
(\bm{q})]^{2}$ with the on-shell energies $\omega _{j} (\bm{q}) \equiv
\sqrt{\bm{q}^{2} + m_{j}^{2}}$ and $\Omega _{j} (\bm{q}) \equiv
\sqrt{\bm{q}^{2} + M_{j}^{2}}$.  This normalization is chosen so that
the expression of the relativistic wave equation~\eqref{eq:KVPsi}
becomes a natural extension of the nonrelativistic \Schr equation (See
Appendix~\ref{sec:B}).  Furthermore, we also introduce the bare state
$\Psi_{a}$, which satisfies the following orthonormal conditions:
\begin{align}
\quad
\langle \Psi_{a}  | \Psi_{b} \rangle 
 = \delta _{a b} ,
\quad 
\langle \bm{q}_{j}^{\text{co}}  | \Psi_{a} \rangle 
=
\langle \Psi_{a}  | \bm{q}_{j}^{\text{co}} \rangle 
=0 .
\label{eq:Psi_norm}
\end{align}
We note that with the normalization~\eqref{eq:qco_norm} and
\eqref{eq:Psi_norm} the complete set of the system is given by
\begin{equation}
1 = 
\sum _{a} | \Psi_{a} \rangle  \langle \Psi_{a}  |
+\sum _{j}
  \int \frac{d^{3} q}{(2 \pi )^{3}}
  \frac{\sqrt{s_{q j}}}
  {2 \omega _{j} (\bm{q}) \Omega _{j} (\bm{q})}
  | \bm{q}_{j}^{\text{co}} \rangle \langle \bm{q}_{j}^{\text{co}} | .
  \label{eq:complete}
\end{equation}

The scattering state $| \bm{q}_{j}^{\rm co} \rangle$ and the bare
state $| \Psi _{a} \rangle$ span the space of the eigenstates of the
kinetic energy operator $\hat{\mathcal{K}}$ which extracts the total
energy squared of the state.  Namely, for the two-body scattering
state $| \bm{q}_{j}^{\text{co}} \rangle$ we have
\begin{equation}
  \hat{\mathcal{K}} | \bm{q}_{j}^{\text{co}} \rangle 
  = s_{qj}| \bm{q}_{j}^{\text{co}} \rangle, \quad
  \langle \bm{q}_{j}^{\text{co}} | \hat{\mathcal{K}}
  = \langle \bm{q}_{j}^{\text{co}} |s_{qj} .
\label{eq:kin_op}
\end{equation}
For the bare state, the eigenvalue of
$\hat{\mathcal{K}}$ is the mass squared of the bare state $\Psi_{a}$,
$M_{a}^{2}$:
\begin{equation}
  \hat{\mathcal{K}}| \Psi_{a} \rangle 
  = M_{a}^{2}| \Psi_{a}  \rangle, \quad
  \langle \Psi_{a}  | \hat{\mathcal{K}}
  = \langle \Psi_{a}  |M_{a}^{2} .
\end{equation}

The dynamics of the system is determined by the interaction operator
$\hat{\mathcal{V}}$.  We again adopt the separable form as
\begin{align}
\langle \bm{q}_{j}^{\prime \, \text{co}} | \hat{\mathcal{V}}
| \bm{q}_{k}^{\text{co}}  \rangle
&=  V_{j k} f_{j} ( q^{\prime \, 2} ) f_{k} ( q^{2} )  ,  %\\
\quad 
\langle \bm{q}_{j}^{\text{co}} |\hat{\mathcal{V}} | \Psi_{a} \rangle 
= \langle \Psi_{a}  | \hat{\mathcal{V}} | \bm{q}_{j}^{\text{co}} \rangle 
=g_{0}^{a, j}f_{j} ( q^{2} ), \quad
\langle \Psi_{a}  | \hat{\mathcal{V}} | \Psi _{b} \rangle = 0,
\label{eq:rel_separable}
\end{align}
where $V_{j k}$ is a real symmetric matrix and $g_{0}^{a, j}$ and
$f_{j}(q^{2})$ are real with an appropriate choice of phases of the
states.\footnote{In relativistic field theory, the coupling $g_{0}^{a,
    j}$ can have an energy dependence from the derivative coupling. We
  do not consider the energy dependence of the coupling, in order to
  ensure a smooth reduction to the results in the previous section in
  the nonrelativistic limit.}  In order to make a three-dimensional
reduction of the scattering equation, we assume that the form factor
$f_{j}(q^{2})$ depends only on the magnitude of the three momentum. We
consider that the wave equation with the operator $\hat{\mathcal{K}} +
\hat{\mathcal{V}}$ contains a resonance $| \Psi ) $ with mass $M_{\rm
  R}$ and width $\Gamma _{\rm R}$ as an
eigenstate~\cite{Mandelzweig:1986hk, Wallace:1989nm}
\begin{equation}
\left [ \hat{\mathcal{K}}  
+ \hat{\mathcal{V}} \right ]
| \Psi ) 
= s_{\rm R}| \Psi ) , 
\quad 
( \Psi | \left [ 
\hat{\mathcal{K}} 
+ \hat{\mathcal{V}} \right ]
= ( \Psi | s_{\rm R}, 
\label{eq:KVPsi}
\end{equation}
where $(\Psi | = \langle \Psi^{*}|$ and $s_{\rm R}=(M_{\rm
  R}-i\Gamma_{\rm R}/2)^{2}$.  By using Eq.~\eqref{eq:complete} we can
decompose the normalization of the resonance vector $(\Psi | \Psi)=1$
as
\begin{equation}
1 = ( \Psi | \Psi ) = \sum _{a} Z_{a} + \sum_{j}X_{j} ,
\label{eq:rel_norm}
\end{equation}
where we have defined the elementariness $Z_{a}$ and compositeness
$X_{j}$ as:
\begin{equation} 
  Z_{a} \equiv (\Psi|\Psi_{a}\rangle\langle\Psi_{a}|\Psi), 
  \quad 
  X_{j} \equiv \int
  \frac{d^{3} q}{(2 \pi )^{3}} \frac{\sqrt{s_{q j}}}{2 \omega
    _{j}(\bm{q}) \Omega _{j}(\bm{q})} \left [ \tilde{\Psi}_{j} ( q )
  \right ] ^{2} ,
\end{equation}
with the momentum space wave function
\begin{equation}
  \langle \bm{q}_{j}^{\text{co}} | \Psi) 
  =\tilde{\Psi}_{j}(q) 
  = ( \Psi |\bm{q}_{j}^{\text{co}} \rangle .
\end{equation}

In the same way with Sec.~\ref{sec:2-3}, the wave function is
determined as
\begin{align}
  \tilde{\Psi}_{j} ( q ) 
  &= \frac{C_{j} f_{j} ( q^{2} )}{s_{\rm R} - s_{q j}}  ,
  \label{eq:WFrel} \\
  C_{j} 
  &= 
  \sum _{k} V_{j k}^{\rm eff} (s_{\rm R})
  \int \frac{d^{3} q^{\prime}}{(2 \pi )^{3}}
  \frac{\sqrt{s_{q^{\prime} k}}} 
  {2 \omega _{k} (\bm{q}^{\prime}) \Omega _{k} (\bm{q}^{\prime \, 2})}
  f_{k}( q^{\prime \, 2} ) \tilde{\Psi} _{k} ( q^{\prime} ) , \\
  V_{j k}^{\rm eff} (s)
  &= V_{j k} + \sum _{a} \frac{g_{0}^{a, j} g_{0}^{a, k}}{s-M_{a}^{2}}
  \label{eq:Veff}
\end{align}
The consistency condition for nonzero $C_{j}$ is given by
\begin{align}
   \det [1-V^{\rm eff}(s_{\rm R})G(s_{\rm R})]=0 ,
   \label{eq:1-VG}
\end{align}
where the loop function $G$ is diagonal with respect to the channel
index and is expressed as
\begin{align}
  G_{j} ( s ) 
  &= 
  \int \frac{d^{3} q}{(2 \pi )^{3}}
  \frac{\sqrt{s_{q j}}}{2 \omega _{j} (\bm{q})
    \Omega _{j} (\bm{q})}
  \frac{[f_{j} ( q^{2} )]^{2}}{s - s_{q j}} %\nonumber \\ &
  = \int \frac{d^{4} q}{(2 \pi )^{4}}
  \frac{i [f_{j} ( q^{2} )]^{2}}{[(P/2 + q)^{2} - m_{j}^{2}]
    [(P/2 - q)^{2} - M_{j}^{2}]} ,
\label{eq:Gloop}
\end{align}
with the energy squared $P^{2} = s$.  The energy squared $s$ in the
denominator of the loop function is considered to have an
infinitesimal positive imaginary part $i \epsilon$: $s \to s + i
\epsilon$.  We note that the dimensional regularization of the loop
function is achieved by setting $f_{j} ( q^{2} ) = 1$ and modifying
the integration variable as $d^{4} k\to \mu _{\rm reg} ^{4-d} d^{d} k$
with the regularization scale $\mu _{\rm reg}$.

In Appendix~\ref{sec:B} we confirm that the wave
equation~\eqref{eq:KVPsi} indeed describes a two-body system governed
by the relativistic scattering equation.  Namely, with the
energy-dependent two-body interaction $V_{j k}^{\rm eff}
(s)$~\eqref{eq:Veff} and the loop function
$G_{j}(s)$~\eqref{eq:Gloop}, the scattering amplitude $T_{j k}(s)$ can
be calculated as
\begin{equation}
T_{j k} ( s ) 
= V_{j k}^{\rm eff} ( s )
+ \sum _{l} V_{j l}^{\rm eff} ( s ) G_{l} (s) T_{l k} ( s ) .
\end{equation}
Therefore, Eq.~\eqref{eq:1-VG} ensures that the scattering amplitude
$T_{j k}(s)$ has a pole at $s=s_{\rm R}$.

\begin{figure}
  \centering
  \begin{tabular*}{\textwidth}{@{\extracolsep{\fill}}cc}
      \PsfigII{0.22}{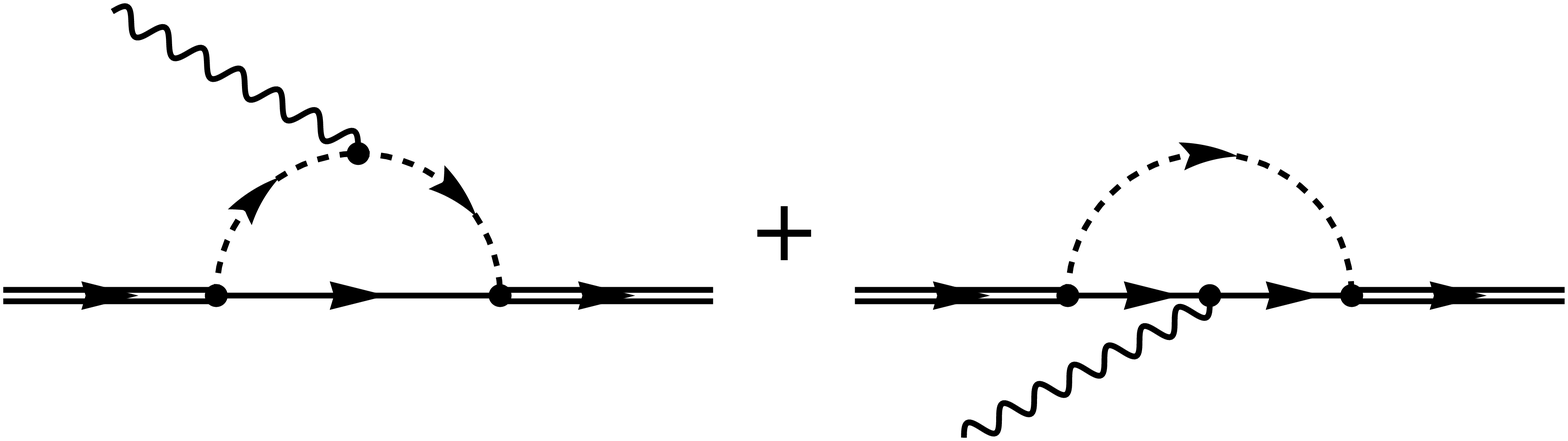} & 
      \PsfigII{0.22}{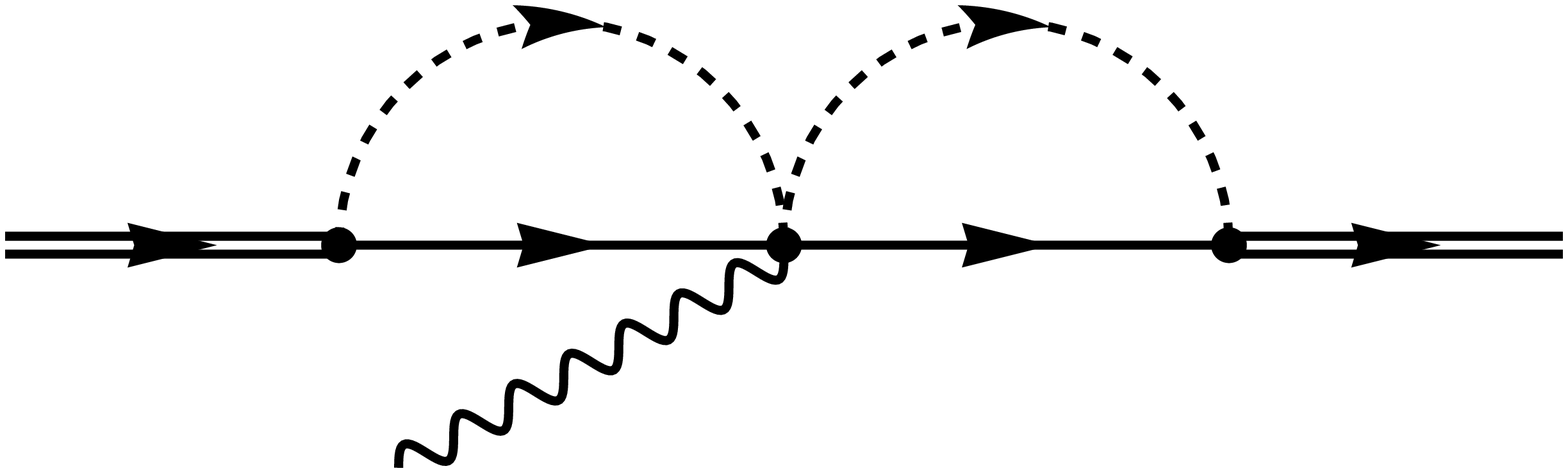} \\
      (a) & (b)
    \end{tabular*}
    \caption{Diagrammatic interpretation of the
      compositeness $X$ (a) and elementariness $Z$ (b). 
      The double and wiggly lines represent the resonance state and
      the probe current, respectively, and the solid and dashed lines
      correspond to the constituent particles.}
  \label{fig:FF}
\end{figure}

By comparing the residue of the resonance pole as in
Eq.~\eqref{eq:resonancecoupling}, we find $C_{j}=g_{j}$, where
\begin{equation}
g_{j} g_{k} = \lim _{s \to s_{\rm R}} 
( s - s_{\rm R} ) T_{j k} (s) , 
\end{equation}
Then, in the same way with Sec.~\ref{sec:2-3}, we obtain
\begin{equation}
\langle \Psi_{a}| \Psi )
= ( \Psi | \Psi_{a} \rangle
=  \sum_{j} \frac{g_{j} g_{0}^{a,j}}{s_{\rm R} - M_{a}^{2}} 
G_{j} ( s_{\rm R} ) . 
\end{equation}
Therefore, we obtain the compositeness and elementariness as
\begin{equation}
  X_{j} = - g_{j}^{2} 
  \left[
  \frac{d G_{j}}{d s} \right]_{s = s_{\rm R}} ,
\label{eq:rel_X}
\end{equation}
\begin{equation}
  Z \equiv \sum _{a} Z_{a} = - \sum _{j,k} g_{k} g_{j} 
  \left [ G_{j} \frac{d V^{\rm eff}_{j k}}{d s} G_{k} 
  \right ] _{s = s_{\rm R}} ,
\label{eq:rel_Z}
\end{equation}
and the sum rule is derived from the normalization~\eqref{eq:rel_norm}
as
\begin{equation}
- \sum _{j, k} g_{k}g_{j}  \left [ 
\delta _{j k} \frac{d G_{j}}{d s}
+G_{j} \frac{d V_{j k}^{\rm eff}}{d s} G_{k} 
\right ] _{s = s_{\rm R}} = 1 .
\label{eq:Widentity}
\end{equation}
This is another derivation of the generalized Ward identity in
Ref.~\cite{Sekihara:2010uz}. In Ref.~\cite{Sekihara:2010uz},
Eq.~\eqref{eq:Widentity} is obtained by attaching one probe current to
the meson-baryon scattering amplitude. The derivative of the loop
function corresponds to the diagrams in Fig.~\ref{fig:FF} (a) in the
soft limit of the probe current. It is therefore consistent to
interpret the first term of Eq.~\eqref{eq:Widentity} as compositeness
which reflects the contribution from the two-body molecule
component. On the other hand, the derivative of the contact
interaction corresponds to the attachment of the probe current to the
interaction vertex [Fig.~\ref{fig:FF} (b)], which represents something
other than the compositeness and thus is understood as the
elementariness.

We note that, although both the compositeness $X_{j}$ and
elementariness $Z$ are complex for resonances, their sum should be
unity, provided that the proper normalization of the wave function is
adopted.  As in the nonrelativistic case, the compositeness
(elementariness) is expressed with the derivative of the loop function
(interaction), and they can be determined by the local behavior of the
interaction and loop function.  Finally we mention that the expression
of the elementariness $Z$ in Eq.~\eqref{eq:rel_Z} coincides with that
derived by matching with the Yukawa theory in
Ref.~\cite{Hyodo:2011qc}.  In this work, we derive $Z$ and $X_{j}$
without specifying the explicit form of the vertex and relate them
with the wave function of the bound and resonance states.

\section{Applications --- structure of dynamically generated
  hadrons}
\label{sec:3}

\subsection{Compositeness and elementariness in chiral dynamics}
\label{sec:3-1}

Having established the compositeness and elementariness in
Eqs.~\eqref{eq:rel_X} and \eqref{eq:rel_Z}, we now turn to the
analysis of physical hadronic resonances by theoretical models with
hadronic degrees of freedom.  One of the most prominent models is the
coupled-channel approach with the chiral perturbation theory.  In this
model the nonperturbative summation of the chiral interaction makes it
possible to generate hadronic resonances dynamically, and hence these
hadronic resonances are often called dynamically generated
hadrons. This framework has been successfully applied to the
description of the low energy hadron scatterings with resonance
states. Among others, the $\Lambda(1405)$ resonance in the strangeness
$S=-1$ meson-baryon scattering~\cite{Kaiser:1995eg, Oset:1997it,
  Oller:2000fj,Lutz:2001yb, Jido:2003cb, Hyodo:2007jq, Ikeda:2011pi,
  Ikeda:2012au, Hyodo:2011ur} and the lightest scalar and vector
mesons in the meson-meson scattering~\cite{Oller:1997ti, Oller:1997ng,
  Oller:1998hw, Oller:1998zr,Truong:1988zp, Truong:1991gv,
  Dobado:1989qm, Dobado:1993ha, Dobado:1996ps, GomezNicola:2001as}
have been extensively studied in this approach.

The compositeness and elementariness have been evaluated in the chiral
model with the simple leading order chiral interaction for
$\Lambda(1405)$ and the scalar mesons in
Ref.~\cite{Sekihara:2012xp}. The compositeness of the $\rho (770)$
meson~\cite{Aceti:2012dd} and $K^{*} (892)$~\cite{Xiao:2012vv} are
also studied in phenomenological models. Here we aim at more
quantitative discussion by using refined chiral models constrained by
the recent experimental data. For this purpose, we employ the
next-to-leading order calculations for
$\Lambda(1405)$~\cite{Ikeda:2011pi, Ikeda:2012au} and for scalar and
vector mesons~\cite{GomezNicola:2001as}.

As we will show below, the scattering amplitudes in
Refs.~\cite{Ikeda:2011pi, Ikeda:2012au, GomezNicola:2001as} can be
reduced to the form of the coupled-channel algebraic equation
\begin{equation}
T_{j k} ( s ) 
= V_{j k} ( s )
+ \sum _{l} V_{j l} ( s ) G_{l} ( s ) T_{l k} ( s ) .
\label{eq:BSeq}
\end{equation}
Here the separable interaction kernel $V_{j k}$ is a symmetric matrix
with respect to the channel indices and depends on the Mandelstam
variable $s$, and $G_{j}$ is the two-body loop function. The explicit
forms of $V_{j k}$ and $G_{j}$ will be given for each model. The
resonances are identified by the poles of the scattering amplitude
$T_{j k}$, and the scattering amplitude can be written in the vicinity
of one of the resonance poles as:
\begin{equation}
T_{j k} ( s ) = \frac{g_{j} g_{k}}{s - s_{\rm R}} 
+ T_{j k}^{\rm BG} ( s ) , 
\label{eq:amp_pole}
\end{equation}
where $g_{j}$ and $s_{\rm R}$ are the coupling constant and the pole
position for the resonance, respectively, and $T_{j k}^{\rm BG}$ is a
background term which is regular at $s \to s_{\rm R}$.

In this study, we utilize the set of the one- and two-body states
introduced in Sec.~\ref{sec:2} as the basis to interpret the structure
of the hadronic resonances in the coupled-channel chiral model.  On
the assumption that the energy dependence of the interaction
originates from channels which do not appear as explicit degrees of
freedom, it has been shown that the final expression of the
compositeness is given by Eq.~\eqref{eq:rel_X} only with the
quantities at the pole position.  Namely, the $j$-channel
compositeness is expressed with the pole position and the residue of
the amplitude and the derivative of the loop function $G_{j}$, which
is obtained with the two-body eigenstates of the free Hamiltonian
$\hat{H}_{0}$:
\begin{equation}
  X_{j} 
  = - g_{j}^{2} \left [ \frac{d G_{j}}{d s} \right ] _{s = s_{\rm R}} . 
\label{eq:chiral_comp}
\end{equation}
On the other hand, the elementariness $Z$ is given by the rest 
of the component out of unity:
\begin{equation}
  Z = 1 - \sum _{j} X_{j} .
\label{eq:chiral_sum-rule}
\end{equation}
By using the interaction $V$ in the coupled-channel
equation~\eqref{eq:BSeq}, the elementariness $Z$ is also given as
\begin{equation}
Z = - \sum _{j,k} g_{k} g_{j} \left [ G_{j} \frac{d V_{j k}}{d s} 
G_{k} \right ] _{s = s_{\rm R}} , 
\label{eq:chiral_elem}
\end{equation}
which measures the contributions from one-body bare states and
implicit two-body states on the basis in Sec.~\ref{sec:2}.  Since the
contribution of the bare state with a large mass gives interaction
with polynomial energy dependence, we are allowed to apply
Eq.~\eqref{eq:chiral_elem} for $V$ with general energy dependence,
which can be reproduced with suitable bare states.

Let us summarize the interpretation of the compositeness and
elementariness for resonances. As shown in Sec.~\ref{sec:2-3}, $X_{j}$
and $Z$ for resonances are in general complex. This fact spoils the
probabilistic interpretation in a strict sense. It is however possible
to interpret the structure of the resonance when one of the real parts
of $X_{j}$ or $Z$ is close to unity and all the other numbers have
small absolute values. In this case, we interpret that the resonance
is dominated by the $j$-th channel component or something other than
the two-body channels involved, respectively, on the basis of the
similarity of the wave function of the stable bound state.

\subsection{Structure of $\Lambda (1405)$}
\label{sec:3-2}

In Refs.~\cite{Ikeda:2011pi, Ikeda:2012au} the low-energy meson-baryon
interaction in the strangeness $S=-1$ sector has been constructed in
chiral perturbation theory up to the next-to-leading order, which
consists of the Weinberg-Tomozawa contact term, the $s$- and
$u$-channel Born terms, and the next-to-leading order contact
terms. After the $s$-wave projection, the interaction kernel $V_{j k}$
depends only on the Mandelstam variable $s$ as a real symmetric
separable interaction.  The explicit form of $V_{j k}$ can be found in
Refs.~\cite{Hyodo:2011ur,Ikeda:2012au}.  The loop function is
regularized by the dimensional regularization:
\begin{align}
G_{j} (s) = & i \mu_{\rm reg}^{4-d} \int \frac{d^{d} q}{(2 \pi )^{d}}
  \frac{1}{[(P/2 + q)^{2} - m_{j}^{2}]
    [(P/2 - q)^{2} - M_{j}^{2}]} 
\notag \\
= & a_{j} ( \mu _{\rm reg} ) 
+ \frac{1}{16 \pi ^{2}} \left [ - 1 
+ \ln \left ( \frac{m_{j}^{2}}{\mu _{\rm reg}^{2}} \right ) 
+ \frac{s + M_{j}^{2} - m_{j}^{2}}{2 s} 
\ln \left ( \frac{M_{j}^{2}}{m_{j}^{2}} \right )
\right . \notag \\ & \left . 
\quad \quad \quad \quad \quad \quad \quad \quad 
- \frac{\lambda ^{1/2} (s, \, m_{j}^{2}, \, M_{j}^{2})}{s} 
\text{artanh} 
\left ( \frac{\lambda ^{1/2} (s, \, m_{j}^{2}, \, M_{j}^{2})}
{m_{j}^{2} + M_{j}^{2} - s} \right ) 
\right ] ,
\label{eq:Gdim_explicit}
\end{align}
with the \Kaellen function $\lambda (x, \, y, \, z) = x^{2} + y^{2} +
z^{2} - 2 x y - 2 y z - 2 z x$.  The finite part is specified by the
subtraction constant $a_{j}(\mu _{\rm reg})$ at the regularization
scale $\mu _{\rm reg}$. Because the meson-baryon one loop is counted
as next-to-next-to-leading order in the baryon chiral perturbation
theory, the amplitude is not renormalizable and hence it depends on
the subtraction constants in this framework. The low-energy constants
in the next-to-leading order contact interaction terms and the
subtraction constants of the loop function have been determined by
fitting to the low-energy total cross sections of $K^{-}p$ scattering
to elastic and inelastic channels, the threshold branching ratios, and
the recent measurement of the $1s$ shift and width of the kaonic
hydrogen~\cite{Bazzi:2011zj, Bazzi:2012eq}. In this approach, the
$\Lambda (1405)$ resonance is associated with two poles of the
scattering amplitude in the complex energy
plane~\cite{Jido:2003cb}. For convenience we refer to the pole which
has higher (lower) mass $M_{\rm R} = \text{Re} \left ( \sqrt{s_{\rm
      R}} \right )$ as the higher (lower) pole. It is expected from
the structure of the Weinberg-Tomozawa interaction that the higher
pole originates in a bound state caused by the $\bar{K} N$
attraction~\cite{Hyodo:2007jq}.

\begin{table}[!t]
  \caption{Compositeness $X_{j}$ and elementariness $Z$ of 
    $\Lambda (1405)$ in the isospin basis.}
  \label{tab:Lambda}
  \centering
  \begin{tabular}{lcc}
    \hline
    & $\Lambda (1405)$, higher pole & $\Lambda (1405)$, lower pole \\
    \hline
    $\sqrt{s_{\rm R}}$ [MeV] 
    & $1424 - 26 i$ & $1381 - 81 i$ \\
    $X_{\bar{K} N}$ 
    & $\phantom{+}1.14 + 0.01 i$ & $-0.39 - 0.07 i$ \\
    $X_{\pi \Sigma}$ 
    & $-0.19 - 0.22 i$ & $\phantom{+}0.66 + 0.52 i$ \\
    $X_{\eta \Lambda}$ 
    & $\phantom{+}0.13 + 0.02 i$ & $-0.04 + 0.01 i$ \\
    $X_{K \Xi}$ 
    & $\phantom{+}0.00 + 0.00 i$ & $-0.00 + 0.00 i$ \\
    $Z$ 
    & $-0.08 + 0.19 i$ & $\phantom{+}0.77 - 0.46 i$ \\
    \hline
  \end{tabular}
\end{table}

With the formulae in Sec.~\ref{sec:3-1} we calculate the pole
positions, compositeness $X_{j}$ and the elementariness $Z$ of the
$\Lambda (1405)$ resonance in this model. Results are summarized in
Table~\ref{tab:Lambda}.  In Refs.~\cite{Ikeda:2011pi, Ikeda:2012au},
the isospin symmetry is slightly broken by the physical hadron masses.
Therefore, we evaluate the compositeness in the charge basis and
define the compositeness in the isospin basis by summing up all the
channels in the charge basis, i.e., $X_{\bar{K} N} = X_{K^{-} p} +
X_{\bar{K}^{0} n}$, and so on. Although there are nonzero
contributions from the $I=1$ channels, $X_{\pi ^{0} \Lambda}$ and
$X_{\eta \Sigma ^{0}}$, to the total normalization, these are
negligible and hence not listed in Table~\ref{tab:Lambda}.  It is
remarkable that the real part of the $X_{\bar{K}N}$ component of the
higher $\Lambda (1405)$ pole is close to unity and its imaginary part
is very small. In addition, the magnitude of the real and imaginary parts
of all the other components is also small ($\lesssim 0.2$). This
indicates that the wave function of the higher $\Lambda (1405)$ pole
is similar to that of the pure $\bar{K}N$ bound state which has
$X_{\bar{K}N}=1$, $X_{i}=0 ~ (i\neq \bar{K}N)$ and $Z=0$. It is
therefore natural to interpret that the higher $\Lambda (1405)$ pole
is dominated by the $\bar{K}N$ composite component. This is consistent
with the non-$qqq$ nature of this pole from the $N_{c}$ scaling
analysis~\cite{Hyodo:2007np,Roca:2008kr}.

On the other hand, for the lower pole, there is a certain amount of
cancellation ($\sim 0.4$) in the real part of the sum
rule~\eqref{eq:Widentity}, and the absolute values of the imaginary
parts are as large as $\sim 0.5$.  Although one may observe relatively
large contributions in $X_{\pi\Sigma}$ and $Z$, the dominance of these
components is comparable with the magnitude of the imaginary
part. Therefore, it is not possible to clearly conclude the structure
of the lower pole from the present analysis.

The compositeness and elementariness of $\Lambda (1405)$ were
calculated in Ref.~\cite{Sekihara:2012xp} using the simple chiral
model with the leading order Weinberg-Tomozawa interaction. The
qualitative features of $X_{j}$ and $Z$ are not changed very much, so
we confirm the earlier results in the present refined model. At the
quantitative level, the results of the lower pole show relatively
larger model dependence. This model dependence also implies the
difficulty of the clear interpretation of the structure of the lower
pole.

Before closing this subsection, we mention that the structure of
$\Lambda (1405)$ was investigated in the complex scaling method in
Refs.~\cite{Dote:2012bu, Dote:2014ema}.  In a $\bar{K} N$-$\pi \Sigma$
two-channel model, the norm of each component is evaluated from the
wave function. It is found that the norm of the $\bar{K} N$ component
of the higher $\Lambda (1405)$ pole is close to unity with a small
imaginary part.  Thus, the result for the higher pole is qualitatively
consistent with ours. On the other hand, the result for the lower pole
in Ref.~\cite{Dote:2014ema} shows the dominance of the $\pi \Sigma$
component.  This is because the complete set to decompose the
resonance wave function in Ref.~\cite{Dote:2014ema} does not contain
the elementary component.  Namely, the application of our formula to
their amplitude would indicate a certain amount of the elementary
component $Z$, as we have found here, since the interaction in
Ref.~\cite{Dote:2014ema} has an energy dependence.  In fact, this is
in accordance with the observation in Ref.~\cite{Dote:2014ema} that
the lower pole disappears when the energy dependence of the
interaction is switched off.

\subsection{Structure of the lightest scalar and vector mesons}
\label{sec:3-3}

The lowest lying scalar and vector mesons in the meson-meson
scattering have been studied in Ref.~\cite{GomezNicola:2001as} using
the Inverse Amplitude Method (IAM) with the chiral interaction up to
the next-to-leading order. The scattering amplitude in the
coupled-channel IAM is given by\begin{equation} T (s) = T_{2} (s)
  \left [ T_{2} (s) - T_{4} (s) \right ] ^{-1} T_{2} (s) ,
\label{eq:IAM}
\end{equation}
where $T_{2}$ and $T_{4}$ are respectively the leading and
next-to-leading order amplitudes in the matrix form with channel
indices from chiral perturbation theory and have been projected to the
orbital angular momentum $L=0$ (scalar) and $L=1$ (vector). In
contrast to the model in the previous subsection, meson-meson one loop
is in the next-to-leading order, and hence the amplitude does not
depend on the renormalization scale. Therefore, the parameters in this
model are the renormalized low energy constants in the next-to-leading
order chiral Lagrangians. These constants are determined by fitting
the experimental meson-meson scattering data such as the $\pi \pi$
scattering up to $\sqrt{s}=1.2 \gev$~\cite{GomezNicola:2001as}. The
lightest scalar mesons $\sigma$, $f_{0}(980)$, and $\kappa$ are found
as poles of the $s$-wave amplitude, while the $a_{0}(980)$ resonance
appears as a cusp at the $K \bar{K}$ threshold but the corresponding
resonance pole is not found. The vector mesons $\rho (770)$ and
$K^{*}(892)$ are also dynamically generated as poles of the $p$-wave
amplitude.

To evaluate the compositeness and elementariness, we rewrite the
amplitude~\eqref{eq:IAM} in the form of Eq.~\eqref{eq:BSeq}. To this
end, we first notice that $T_{4}$ can be decomposed as the $s$-channel
loop part and the rest:
\begin{equation}
T_{4} = T_{2} G T_{2}+T_{4\text{,non-}G}  , 
\label{eq:T4_decomp}
\end{equation}
where $T_{4\text{,non-}G} (s)$ consists of the next-to-leading order
tree-level amplitudes, tadpoles, and $t$- and $u$-channel loop
contributions. The loop function $G_{j} (s)$ in
Eq.~\eqref{eq:T4_decomp} is given by
\begin{align}
G_{j} (s) = & 
\frac{1}{16 \pi ^{2}} \left [ - 1 
+ \left ( \frac{M_{j}^{2} - m_{j}^{2}}{2 s} 
+ \frac{m_{j}^{2} + M_{j}^{2}}{2 ( m_{j}^{2} - M_{j}^{2} )} \right )
\ln \left ( \frac{M_{j}^{2}}{m_{j}^{2}} \right )
\right . \notag \\ & \left . 
\quad \quad \quad 
- \frac{\lambda ^{1/2} (s, \, m_{j}^{2}, \, M_{j}^{2})}{s} 
\text{artanh} 
\left ( \frac{\lambda ^{1/2} (s, \, m_{j}^{2}, \, M_{j}^{2})}
{m_{j}^{2} + M_{j}^{2} - s} \right ) 
\right ] .
\label{eq:G_IAM}
\end{align}
Note that there is no degrees of freedom of the subtraction constant;
the finite part is determined by the low energy constants included in
$T_{4\text{,non-}G}$. We then define
\begin{equation}
V
\equiv T_{2} ( T_{2} - T_{4\text{,non-}G} )^{-1} T_{2} .
\label{eq:V_IAM}
\end{equation}
It is easily checked that the amplitude in IAM~\eqref{eq:IAM} is
formally equivalent to Eq.~\eqref{eq:BSeq} with the
interaction~\eqref{eq:V_IAM} and the loop
function~\eqref{eq:G_IAM}. We thus interpret Eq.~\eqref{eq:V_IAM} as
the effective interaction kernel used in IAM.  Physically, this
interaction kernel contains not only the chiral interaction up to the
next-to-leading order but also the nonperturbative summation of
contributions from $t$- and $u$-channel loops.  We also note that the
interaction kernel~\eqref{eq:V_IAM} can have a nonzero imaginary part
due to the contributions from the $t$- and $u$-channel loops, which
will disappear in the nonrelativistic limit.

Before evaluating the compositeness, let us focus on the structure of
the interaction kernel~\eqref{eq:V_IAM}. Because of the $ ( T_{2} -
T_{4\text{,non-}G} )^{-1} $ factor, the interaction kernel can have a
pole when $\det[T_{2}(s)- T_{4\text{,non-}G}(s)]=0$ is
satisfied. Thus, even though the IAM is constructed from chiral
perturbation theory without bare fields of the scalar and vector
mesons, there can be a pole contribution in the effective interaction
$V$. In fact, we find poles in the vector-channel $V$ near the
physical resonances as
\begin{equation}
   \rho \text{ channel}: 746 - 11 i  \mev ,\quad 
   K^{\ast} \text{ channel}: 890 - 0 i \mev .
   \label{eq:barepole}
\end{equation}
In contrast, in the scalar channel there is no pole contribution in
the relevant energy region. The pole structure of the interaction $V$
can be related to the origin of resonances in the full amplitude $T$.

\begin{table}[!t]
  \caption{Compositeness $X_{j}$ and elementariness $Z$ of scalar mesons 
    in the isospin basis.}
  \label{tab:scalar}
  \centering
  \begin{tabular}{lccc}
    \hline
    & $f_{0}(500) = \sigma$ & $f_{0}(980)$ & 
    $K_{0}^{\ast} (800) = \kappa$ \\
    \hline
    $\sqrt{s_{\rm R}}$ [MeV] & $443 - 217 i$ & $988 - 4 i$ & 
    $750 - 227 i$ \\
    $X_{\pi \pi}$   & $-0.09 + 0.37 i$ & $\phantom{+}0.00 - 0.00 i$ & 
    --- \\
    $X_{K \bar{K}}$ & $-0.01 - 0.00 i$ & $\phantom{+}0.87 - 0.04 i$ &
    --- \\
    $X_{\eta \eta}$ & $-0.00 + 0.00 i$ & $\phantom{+}0.06 + 0.01 i$ & 
    --- \\
    $X_{\pi K}$    & --- & --- & 
    $\phantom{+}0.32 + 0.36 i$ \\
    $X_{\eta K}$   & --- & --- & 
    $-0.01 - 0.00 i$ \\
    $Z$           & $\phantom{+}1.09 - 0.37 i$ & $\phantom{+}0.07 + 0.02 i$ & 
    $\phantom{+}0.70 - 0.36 i$ \\
    \hline
  \end{tabular}
\end{table}

Now let us evaluate the compositeness and elementariness of the
lightest scalar and vector mesons described by the coupled-channel IAM
developed in Ref.~\cite{GomezNicola:2001as} on the basis to of the
one- and two-body states introduced in Sec.~\ref{sec:2}. The obtained
values of the compositeness and elementariness are listed in
Tables~\ref{tab:scalar} (scalar channels) and \ref{tab:vector} (vector
channels).  In the scalar channels, the $f_{0}(980)$ resonance shows a
clear property; the real part of the $K \bar{K}$ component is close to
unity while other components are smaller than $0.07$. This indicates
that the $f_{0}(980)$ resonance is dominated by the $K \bar{K}$
component.  On the other hand, the results for the $\sigma$ and
$\kappa$ resonances are subtle; the largest component seems to be $Z$,
but its imaginary part is not small, $\sim 0.37$. We thus refrain from
interpreting the structure of the $\sigma$ and $\kappa$ resonance from
$X$ and $Z$.  In Refs.~\cite{Pelaez:2003dy,Pelaez:2004xp} the non-$q
\bar{q}$ nature of the scalar mesons is implied from the $N_{c}$
scaling behavior. Our conclusion of the $K \bar{K}$ dominance of
$f_{0}(980)$ is consistent with the $N_{c}$ scaling analysis.

\begin{table}[!t]
  \caption{Compositeness $X_{j}$ and elementariness $Z$ of vector mesons
    in the isospin basis.}
  \label{tab:vector}
  \centering
  \begin{tabular}{lcc}
    \hline
    & $\rho (770)$ & $K^{\ast} (892)$ \\
    \hline
    $\sqrt{s_{\rm R}}$ [MeV] & $760 - 84 i$ & $885 - 22 i$ \\
    $X_{\pi \pi}$   & $-0.08 + 0.03 i$ & --- \\
    $X_{K \bar{K}}$ & $-0.02 + 0.00 i$ & --- \\
    $X_{\pi K}$    & --- & $-0.03 + 0.04 i$ \\
    $X_{\eta K}$   & --- & $-0.03 + 0.00 i$ \\
    $Z$           & $\phantom{+}1.10 - 0.04 i$ & $\phantom{+}1.06 - 0.04 i$ \\
    \hline
  \end{tabular}
\end{table}

In the vector channels\footnote{In the framework of IAM, the loop
  function in $p$ wave is identical to that in $s$ wave. On the other
  hand, with a nonrelativistic separable interaction, the loop
  function in the $l$-th partial wave should contain the $q^{2l}$
  factor in the integrand~\cite{Aceti:2012dd, Xiao:2012vv}. This is to
  ensure the correct low energy behavior of the amplitude
  $\mathcal{F}_{l}(q) \sim q^{2l}$. The difference of the loop
  function may be regarded as the difference of the definition of $Z$
  and $X_{j}$ (basis to form the complete set). We however note that
  the present definition leads to $Z = 0$ in the $B \to 0$ limit even
  in $p$ wave, while the definition in Refs.~\cite{Aceti:2012dd,
    Xiao:2012vv} does not constrain the value of $Z$ at threshold for
  nonzero $l$. The general threshold behavior is consistent with the
  latter~\cite{Hyodo:2014bda}, so the present definition would lead to
  a special behavior near the threshold.  In practice, the $\rho
  (770)$ and $K^{\ast} (892)$ mesons locate away from the threshold
  energies of meson-meson channels, so the special nature of the
  definition would not cause a problem in the present analysis.}, we
find that, for both the $\rho (770)$ and $K^{*} (892)$ mesons, the
real part of the elementariness $Z$ is close to unity and the
magnitude of the imaginary part is less than $0.1$.  This indicates
that the structure originates in the elementary component.  This is
consistent with the finding of the pole contribution in the
interaction kernel $V$ for the vector channels. In fact, the physical
pole position in Table~\ref{tab:vector} is very close to that in the
effective interaction~\eqref{eq:barepole}. We thus conclude that these
vector mesons are not dominated by the two-meson composite
structure. This is consistent with the $N_{c}$ scaling analysis in
Refs.~\cite{Pelaez:2003dy,Pelaez:2004xp}, which indicates the $q
\bar{q}$ structure of vector mesons.

The compositeness of scalar mesons [$\sigma$, $f_{0}(980)$, and
$a_{0}(980)$] has been studied in Ref.~\cite{Sekihara:2012xp} using
the leading order chiral interaction. The qualitative tendency of the
results of the $\sigma$ and $f_{0}(980)$ is similar with the present
calculation, while the dominance of the $K \bar{K}$ component of
$f_{0}(980)$ is much clear in the present results.  Also for the
vector mesons, the present calculation in IAM with the next-to-leading
order chiral interaction is consistent with the previous
phenomenological ones in Refs.~\cite{Aceti:2012dd, Xiao:2012vv}, which
suggest that $\rho (770)$ and $K^{\ast} (892)$ are elementary.

\subsection{Structure of other hadrons}
\label{sec:3-4}

In the preceding subsections we have evaluated the compositeness and
elementariness of $\Lambda (1405)$, light scalar mesons, and light
vector mesons using the scattering amplitudes calculated in chiral
dynamics with systematic improvements by higher order contributions.
In this subsection we also discuss the compositeness and
elementariness of $N (1535)$ and $\Lambda (1670)$ in a simplified
model with the lowest order Weinberg-Tomozawa interaction.  Although a
systematic analysis is not performed for these resonances, the model
with appropriate subtraction constants~\cite{Inoue:2001ip,
  Oset:2001cn} describes $N (1535)$ and $\Lambda (1670)$ reasonably
well.

\begin{table}
  \caption{Compositeness $X_{j}$ and elementariness $Z$ of $N (1535)$ 
    and $\Lambda (1670)$ in the isospin basis.}
  \label{tab:other}
  \centering
  \begin{tabular}{lc|lc}
    \hline
    \multicolumn{2}{c|}{$N (1535)$} & \multicolumn{2}{c}{$\Lambda (1670)$} \\
    \hline
    $\sqrt{s_{\rm R}}$ [MeV] & 
    $1529 - 37 i$ & 
    $\sqrt{s_{\rm R}}$ [MeV] & 
    $1678 - 21 i$ \\
    $X_{\pi N}$     & $-0.02 - 0.01 i$ & 
    $X_{\bar{K} N}$       & $ \phantom{-} 0.03 + 0.00 i $ \\
    $X_{\eta N}$    & $-0.03 + 0.23 i$ &
    $X_{\pi \Sigma}$   & $ \phantom{-} 0.00 + 0.00 i $ \\
    $X_{K \Lambda}$ & $ \phantom{-} 0.09 - 0.04 i $ &
    $X_{\eta \Lambda}$ & $-0.09 + 0.16 i$ \\
    $X_{K \Sigma}$  & $ \phantom{-} 0.26 - 0.09 i $ &
    $X_{K \Xi}$        & $ \phantom{-} 0.53 - 0.10 i $ \\
    $Z$         & $ \phantom{-} 0.70 - 0.09 i $ &
    $Z$            & $ \phantom{-} 0.53 - 0.06 i $ \\
    \hline
  \end{tabular}
\end{table}

Using $V(s)$ and $G(s)$ with the subtraction parameters given in
Ref.~\cite{Inoue:2001ip} for $N (1535)$ and Ref.~\cite{Oset:2001cn}
for $\Lambda (1670)$, we calculate the compositeness and
elementariness of $N (1535)$ and $\Lambda (1670)$.  The results are
listed in Table~\ref{tab:other}.  First of all, interestingly, for
both resonances the imaginary parts of the values of the compositeness
$X_{j}$ and elementariness $Z$ are relatively small. This may allow us
to interpret $X_{j}$ and $Z$ as the components of the resonance
state. For $N (1535)$, $Z$ is a dominant piece with a relatively small
imaginary part. This suggests that $N (1535)$ in the present model has
a large component originating from contribution other than the
pseudoscalar meson-baryon dynamics considered, in accordance with
Ref.~\cite{Hyodo:2008xr}. In contrast, for $\Lambda (1670)$ the $K
\Xi$ compositeness $X_{K \Xi}$ and the elementariness $Z$ share unity
half-and-half. This implies that in the present model the $K \Xi$
composite state plays a substantial role for the $\Lambda (1670)$ pole
together with a bare state coming from components other than
meson-baryon systems.  This conclusion on $\Lambda (1670)$ is
consistent with the discussion with the natural renormalization scheme
in Ref.~\cite{Doring:2010rd}.

Here we emphasize that both $N (1535)$ and $\Lambda (1670)$ discussed
in this subsection are described by scattering amplitudes which do not
fully reproduce the experimental data in relevant
energies~\cite{Hyodo:2002pk, Hyodo:2003qa}. For more realistic
discussion, it is desirable to improve the theoretical models so as to
reproduce the experimental data well, for instance, by taking into
account the interplay between $N (1535)$ and $N
(1650)$~\cite{Nieves:2001wt}, by including the vector meson-baryon
channels~\cite{Garzon:2014ida}, and by implementing higher order
terms.

\section{Conclusion}
\label{sec:4}

In this study we have developed a framework to investigate the
internal structure of bound and resonance states with their
compositeness and elementariness by using their wave functions.  For
this purpose we have explicitly taken into account both one-body bare
states and two-body scattering states as the basis to interpret the
structure of bound and resonance states.  Compositeness and
elementariness are respectively defined as the contributions from the
two-body scattering states and the one-body bare states to the
normalization of the total wave function.  After reviewing the
formulation for the bound state, we have discussed the extension to
the resonance state.

Because the wave function is analytically obtained for a separable
interaction, we have explicitly written down the wave function for a
bound state in a general separable interaction and obtained the
expressions of the compositeness and elementariness.  We have
demonstrated that the compositeness is determined by the residue of
the scattering amplitude and the energy dependence of the loop
function at the pole position.  Therefore, once one has the loop
function, which is the Green function of the free two-body
Hamiltonian, one can obtain the compositeness only from the bound
state properties.  On the other hand, we have found that the
elementariness is obtained with the energy dependence of the effective
two-body interaction.  It is an interesting finding that the energy
dependence of the two-body effective interaction arises from implicit
channels which do not appear as explicit degrees of freedom but are
effectively taken into account for the two-body interaction in the
practical model space.  These implicit channels contain the two-body
scattering states as well as the one-body bare states.  We have also
shown the sum rule of the compositeness and elementariness.  We have
proved that, with multiple bare states, the formulae of the
compositeness and elementariness can be applied to interactions with
an arbitrary energy dependence.
Of particular value is the derivation of the Weinberg's relation for
the scattering length and effective range in the weak binding limit.
In the present formulation, thanks to the separable interaction, the
scattering amplitude is analytically obtained.  With this fact we have
explicitly performed the expansion of the amplitude around the
threshold to derive the Weinberg's relation.  In this derivation, the
higher order corrections come from the explicit expression of the form
factor as well as higher order derivatives in the expansion. The
limitation of the formula due to the existence of the CDD pole is
clearly linked to the breakdown of the effective range expansion.

Our discussion on the wave function has been extended to resonance
states with the Gamow vectors.  The use of the Gamow vector enables us
to have finite normalization of the resonance wave function.  For a
resonance state, by definition both the compositeness and
elementariness become complex, which are difficult to interpret.
Nevertheless, utilizing the fact that the compositeness and
elementariness are defined by the wave functions, we have proposed the
interpretation of the structure of certain class of resonance states,
on the basis of the similarity of the wave function of the bound
state.  Namely, if the compositeness in a channel (elementariness) is
close to unity with small imaginary part and all the other components
have small absolute values, this resonance state can be considered to
be a composite state in the channel (an elementary state).  Finally we
have given the expressions of the compositeness and elementariness
with a general separable interaction in a relativistic covariant form
by considering a relativistic scattering with a three-dimensional
reduction.

As applications, the expression of the compositeness in a relativistic
form has been used to investigate internal structure of hadronic
resonances, on the assumption that the energy dependence of the
interaction originates from the implicit channels.  By employing the
chiral coupled-channel scattering models with interactions up to the
next to leading order, we have observed that the higher pole of
$\Lambda (1405)$ and $f_{0}(980)$ are dominated by the $\bar{K} N$ and
$K \bar{K}$ composite states, respectively, while the vector mesons
$\rho (770)$ and $K^{\ast} (892)$ are elementary.

Finally we emphasize that the fact that constituent hadrons are
observable as asymptotic states in QCD is essential to construct the
two-body wave functions and to determine the compositeness for
hadronic resonances.

\section*{Acknowledgments}

The authors greatly acknowledge T.~Myo and A.~Dot\'e for discussions
on the Gamow vectors for resonance states, J.~Nebreda on the
theoretical description of scalar and vector mesons, and H.~Nagahiro
and A.~Hosaka on the physical interpretation of compositeness.
The authors also thank E.~Oset for his careful reading of the
manuscript and stimulating discussions during the stay of T.~S. in
Valencia supported by JSPS Open Partnership Bilateral Joint Research
Projects.
This work is partly supported by the Grant-in-Aid for Scientific
Research from MEXT and JSPS (%
24740152 % Hyodo-san
and 25400254% Jido-san
) and by the Yukawa International Program for Quark-Hadron Sciences
(YIPQS).

\appendix

\section{Proof of Eq.~(62)} % \eqref{eq:ZnoN} 
\label{sec:A}

In this Appendix we prove the relation in Eq.~\eqref{eq:ZnoN}.  In
order to specify the problem, we consider a nonrelativistic stable
bound system described with $N$ two-body channels, in which the $j$-th
channel compositeness $X_{j}$ and the elementariness $Z$ can be
expressed as
\begin{equation}
X_{j} = - g_{j}^{2} 
\left [\frac{d G_{j}}{d E}\right ] _{E = M_{\rm B}} 
\quad 
(j = 1, \, \cdots , \, N ) ,
\label{eqA:Xj}
\end{equation}
\begin{equation}
Z = - \sum _{j , k = 1}^{N}
g_{k}g_{j} \left [ G_{j} \frac{d v^{\rm eff}_{j k}}{d E} 
G_{k} \right ] _{E = M_{\rm B}} ,
\end{equation}
with the coupling constant $g_{j}$, the loop function $G_{j}$, and the
two-body effective interaction $v_{j k}^{\rm eff}$.  Then we make an
implementation of a scattering channel $N$ into the effective
interaction, in the same manner as in~\cite{Hyodo:2007jq}:
\begin{equation}
w_{j k} ( E )
= v_{j k}^{\rm eff} 
+ v_{j N}^{\rm eff} \frac{G_{N} ( E )}{1 - v_{N N}^{\rm eff} G_{N} ( E )}
v_{N k}^{\rm eff} 
\quad 
( j, \, k = 1, \, \cdots , \, N - 1 ) .
\label{eqA:vnoN}
\end{equation}
When we adopt the effective interaction $w_{j k}$ for the $N - 1$
two-body channels, the elementariness $Z^{w}$ may be able to be
calculated by the derivative of the effective interaction $w_{j k}$ as
\begin{equation}
Z^{w} = - \sum _{j , k = 1}^{N - 1} 
g_{k} g_{j} \left [ G_{j} \frac{d w_{j k}}{d E} 
G_{k} \right ] _{E = M_{\rm B}} .
\end{equation}
Now we would like to prove that $Z^{w}$ can be expressed as
\begin{equation}
Z^{w} = Z + X_{N} .
\label{eqA:prove}
\end{equation} 

For this purpose we first note that the coupling constant $g_{j}$
satisfies the following bound state condition:
\begin{equation}
\sum _{k} \left [ \delta _{j k} - v_{j k}^{\rm eff} G_{k} 
\right ] _{E = M_{\rm B}} g_{k} = 0 .
\label{eqA:1-VG}
\end{equation}
In the following equations we omit the argument of the functions $v_{j
  k}^{\rm eff}$, $G_{j}$, and so on, since we always take $E = M_{\rm
  B}$ in this Appendix.  From the condition~\eqref{eqA:1-VG} we can
express $g_{N}$ in terms of other coupling constants $g_{j}$ ($j \ne
N$) as
\begin{equation}
g_{N} = \frac{1}{1 - v_{N N}^{\rm eff} G_{N}} 
\sum _{j = 1}^{N - 1} g_{j} G_{j} v_{j N}^{\rm eff} 
= \frac{1}{1 - v_{N N}^{\rm eff} G_{N}} 
\sum _{k = 1}^{N - 1} g_{k} G_{k} v_{N k}^{\rm eff} .
\label{eqA:gN}
\end{equation}

We prove the relation~\eqref{eqA:prove} by calculating first the
derivative of the effective interaction $w$.  Namely, from
Eq.~\eqref{eqA:vnoN} its derivative can be evaluated as
\begin{align}
\frac{d w_{j k}}{d E}
= & \frac{d v_{j k}^{\rm eff}}{d E} 
+ \frac{d v_{j N}^{\rm eff}}{d E} 
\frac{G_{N}}{1 - v_{N N}^{\rm eff} G_{N}}
v_{N k}^{\rm eff} 
+ v_{j N}^{\rm eff}
\frac{d G_{N}}{d E} \frac{1}{1 - v_{N N}^{\rm eff} G_{N}}
v_{N k}^{\rm eff} 
\notag \\
& + v_{j N}^{\rm eff} 
\frac{G_{N}}{\left ( 1 - v_{N N}^{\rm eff} G_{N} \right )^{2}}
\frac{d \left ( v_{N N}^{\rm eff} G_{N} \right )}{d E} 
v_{N k}^{\rm eff} 
+ v_{j N}^{\rm eff} \frac{G_{N}}{1 - v_{N N}^{\rm eff} G_{N}}
\frac{d v_{N k}^{\rm eff}}{d E} .
\end{align}
Therefore, the elementariness $Z^{w}$ becomes
\begin{align}
Z^{w} = - \sum _{j , k = 1}^{N - 1} 
g_{k} g_{j} & G_{j} G_{k} 
\left [ 
\frac{d v_{j k}^{\rm eff}}{d E} 
+ \frac{d v_{j N}^{\rm eff}}{d E} 
\frac{G_{N}}{1 - v_{N N}^{\rm eff} G_{N}}
v_{N k}^{\rm eff} 
+ v_{j N}^{\rm eff}
\frac{d G_{N}}{d E} \frac{1}{1 - v_{N N}^{\rm eff} G_{N}}
v_{N k}^{\rm eff} 
\right .
\notag \\
& 
\left . + v_{j N}^{\rm eff} 
\frac{G_{N}}{\left ( 1 - v_{N N}^{\rm eff} G_{N} \right )^{2}}
\frac{d \left ( v_{N N}^{\rm eff} G_{N} \right )}{d E} 
v_{N k}^{\rm eff} 
+ v_{j N}^{\rm eff} \frac{G_{N}}{1 - v_{N N}^{\rm eff} G_{N}}
\frac{d v_{N k}^{\rm eff}}{d E} \right ] .
\end{align}
Then using Eq.~\eqref{eqA:gN} we can rewrite $Z^{w}$ as
\begin{align}
Z^{w} = & - \sum _{j , k = 1}^{N - 1} 
g_{k} g_{j} G_{j} G_{k} 
\frac{d v_{j k}^{\rm eff}}{d E} 
- \sum _{j=1}^{N-1} g_{j} G_{j} \frac{d v_{j N}^{\rm eff}}{d E} 
G_{N} g_{N}
- \sum _{j=1}^{N-1} g_{j} G_{j} v_{j N}^{\rm eff}
\frac{d G_{N}}{d E} g_{N}
\notag \\
& 
- g_{N} G_{N}
\frac{d \left ( v_{N N}^{\rm eff} G_{N} \right )}{d E} g_{N}
- \sum _{k=1}^{N-1}
g_{k} G_{k} G_{N} g_{N} \frac{d v_{N k}^{\rm eff}}{d E} .
\end{align}
The third term of the right-hand side can be further translated by
multiplying $1 = (1 - v_{N N}^{\rm eff} G_{N}) / (1 - v_{N N}^{\rm
  eff} G_{N})$ as
\begin{equation}
- \sum _{j=1}^{N-1} g_{j} G_{j} v_{j N}^{\rm eff}
\frac{d G_{N}}{d E} g_{N}
= - g_{N} \left ( 1 - v_{N N}^{\rm eff} G_{N} \right )
\frac{d G_{N}}{d E} g_{N} ,
\end{equation}
which is combined with the fourth term to give
\begin{equation}
- g_{N} \left ( 1 - v_{N N}^{\rm eff} G_{N} \right )
\frac{d G_{N}}{d E} g_{N} 
- g_{N} G_{N}
\frac{d \left ( v_{N N}^{\rm eff} G_{N} \right )}{d E} g_{N}
= - g_{N}^{2} \frac{d G_{N}}{d E} 
- g_{N}^{2} G_{N}^{2} \frac{d v_{N N}^{\rm eff}}{d E} .
\end{equation}
As a consequence, the elementariness $Z^{w}$ becomes
\begin{align}
Z^{w} = & - \sum _{j , k = 1}^{N - 1} 
g_{k} g_{j} G_{j} 
\frac{d v_{j k}^{\rm eff}}{d E} G_{k} 
- \sum _{j=1}^{N-1} g_{j} G_{j} \frac{d v_{j N}^{\rm eff}}{d E} 
G_{N} g_{N}
- g_{N}^{2} \frac{d G_{N}}{d E} 
- g_{N}^{2} G_{N}^{2} \frac{d v_{N N}^{\rm eff}}{d E} 
\notag \\
& - \sum _{k=1}^{N-1}
g_{k} g_{N} G_{N} \frac{d v_{N k}^{\rm eff}}{d E} G_{k} 
\notag \\
= & - \sum _{j , k = 1}^{N} 
g_{k} g_{j} G_{j} \frac{d v_{j k}^{\rm eff}}{d E} G_{k} 
- g_{N}^{2} \frac{d G_{N}}{d E} ,
\end{align}
which completes the proof of Eq.~\eqref{eqA:prove}.  By repeating the
above procedure, one can make an implementation of two or more
two-body channels.  Moreover, in a similar way, one can prove that the
contribution of the bare states can be expressed by the derivative of
the Green function like Eq.~\eqref{eqA:Xj} when the bare states are
not counted into the implicit channels.

\section{Conventions of relativistic two-body state and two-body
  equation}
\label{sec:B}

In this Appendix we summarize our conventions of the two-body state in
the relativistic kinematics and confirm that the wave
equation~\eqref{eq:KVPsi} indeed describes a two-body system whose
motion is governed by the Klein-Gordon equation.  In the following we
concentrate on single-channel kinematics of the two-body system, but
generalization to multi-channel kinematics is straightforward.

\subsection{Normalization of states}

First we consider an on-shell one-body state of a scalar field of mass
$m$ with definite three-dimensional momentum $\bm{p}$, $| \bm{p}
\rangle$, whose normalization is defined as follows:
\begin{equation}
\langle \bm{p}^{\prime} | \bm{p} \rangle
= 2 \sqrt{\bm{p}^{2} + m^{2}} 
(2 \pi )^{3} \delta ^{3} (\bm{p}^{\prime} - \bm{p}) . 
\label{eqA:norm}
\end{equation}
Since we do not explicitly treat spin components of scattering baryons
in this paper, we use the above normalization also for baryons.  

Next we construct a two-body state, in which both two particles are on
the mass shell and the relative momentum is denoted as $\bm{q}$ in the
center-of-mass frame, used in Sec.~\ref{sec:2-4}. In this kinematical
condition, the momenta of two particles are given by $p_{1}^{\mu}
= (\omega (\bm{q}), \, \bm{q})$ and $p_{2}^{\mu} = (\Omega (\bm{q}),
\, -\bm{q})$, where $\omega (\bm{q}) \equiv \sqrt{\bm{q}^{2} + m^{2}}$
and $\Omega (\bm{q}) \equiv \sqrt{\bm{q}^{2} + M^{2}}$ are the
on-shell energies of two particles with $m$ ($M$) being the mass of
the first (second) particle, and the total momentum becomes $P^{\mu}
\equiv p_{1}^{\mu} + p_{2}^{\mu} = (\sqrt{s_{q}}, \, \bm{0})$ with
$\sqrt{s_{q}} \equiv \omega (\bm{q}) + \Omega (\bm{q})$.  Then the
two-body state with relative momentum $\bm{q}$, $| \bm{q}^{\rm co}
\rangle$, can be defined by using product of two one-body states, $|
\bm{p}_{1} \rangle \otimes | \bm{p}_{2} \rangle$.  In this study we
adopt the following normalization of $| \bm{q}^{\rm co} \rangle$:
\begin{equation}
| \bm{q}^{\rm co} \rangle 
\equiv \mathcal{N}_{s q} | \bm{q}_{1} \rangle \otimes | - \bm{q}_{2} \rangle , 
\quad 
\langle \bm{q}^{\rm co} |
\equiv \mathcal{N}_{s q}^{\ast} 
\langle \bm{q}_{1} | \otimes \langle - \bm{q}_{2} | , 
\quad
| \mathcal{N}_{s q} |^{2} \equiv \frac{1}{2 \mathcal{V}_{3} \sqrt{s_{q}}} .
\label{eqA:qco}
\end{equation}
In the normalization factor $\mathcal{N}_{s q}$, $\mathcal{V}_{3}$ is
the total spatial volume and is related to the delta function for the
momentum as $\mathcal{V}_{3} = (2 \pi )^{3} \delta ^{3} (\bm{0})$.
The advantage to adopt this normalization factor is that the
expression of the relativistic two-body wave equation becomes a
natural extension of the nonrelativistic \Schr equation, as we will
see in the next subsection.

With the definition of the two-body state $| \bm{q}^{\rm co} \rangle$
in Eq.~\eqref{eqA:qco} and the normalization of the one-body state in
Eq.~\eqref{eqA:norm}, we can calculate the normalization for $|
\bm{q}^{\rm co} \rangle$ in a straightforward way as
\begin{equation}
\langle \bm{q}^{\prime \, \text{co}} | \bm{q}^{\rm co} \rangle
= \frac{2 \omega (\bm{q}) \Omega (\bm{q})}
{\sqrt{s_{q}}} (2 \pi )^{3} \delta ^{3} (\bm{q}^{\prime} - \bm{q}) . 
\label{eqA:norm_rel}
\end{equation}
This normalization leads to the projection operator to the two-body
state:
\begin{equation}
\hat{\mathcal{P}}_{\rm two} = \int \frac{d^{3} q}{(2 \pi )^{3}}
\frac{\sqrt{s_{q}}}{2 \omega (\bm{q}) \Omega (\bm{q})}
| \bm{q}^{\rm co} \rangle \langle \bm{q}^{\rm co} | ,
\label{eqA:unity_two}
\end{equation} 
which corresponds to a part of the completeness condition.

\subsection{Relativistic wave equation and scattering equation}

Now we would like to confirm that the wave equation~\eqref{eq:KVPsi}
indeed describes a two-body system whose motion is governed by the
Klein-Gordon equation, by deriving the scattering equation from the
operators in the wave equation.  Here, in the same manner as in
Sec.~\ref{sec:2}, we introduce a one-body bare state and a two-body
scattering state, and assume that the bare state contribution is
effectively contained in the two-body interaction $V^{\rm eff}$.  In
this sense, the relation in Eq.~\eqref{eqA:unity_two} coincides with
the completeness condition; $\hat{\mathcal{P}}_{\rm two} = 1$.

In general, the wave equation can be composed of the free two-body
Green's operator $\hat{\mathcal{G}} (s)$ and the two-body interaction
operator $\hat{\mathcal{V}}^{\rm eff} (s)$.  The two-body Green's
operator $\hat{\mathcal{G}} (s)$ is defined as $\hat{\mathcal{G}} (s)
\equiv 1 / (s - \hat{\mathcal{K}})$ with the kinetic energy operator
$\hat{\mathcal{K}}$ so that\footnote{In the nonrelativistic framework
  the two-body Green's operator is $\hat{\mathcal{G}}(E) = 1 / (E -
  \hat{H}_{0})$ with $\hat{H}_{0}$ being the free Hamiltonian and
  Eq.~\eqref{eqA:Psi_eq} is reduced to the \Schr equation.}
\begin{equation}
\hat{\mathcal{G}} (s) | \bm{q}^{\rm co} \rangle 
= \frac{1}{s - s_{q}} | \bm{q}^{\rm co} \rangle , 
\quad 
\langle \bm{q}^{\rm co} | \hat{\mathcal{G}} (s)
= \frac{1}{s - s_{q}} \langle \bm{q}^{\rm co} | . 
\end{equation}
On the other hand, two-body interaction operator
$\hat{\mathcal{V}}^{\rm eff} (s)$ has a general separable interaction
as in Eq.~\eqref{eq:rel_separable}, thus we have
\begin{equation}
\langle \bm{q}^{\prime \, \text{co}} | \hat{\mathcal{V}}^{\rm eff} (s)
| \bm{q}^{\rm co} \rangle
= V^{\rm eff} (s) f ( q^{2} ) f ( q^{\prime \, 2} ) , 
\label{eqA:V}
\end{equation}
where $V^{\rm eff}(s)$ corresponds to the interaction in
Eq.~\eqref{eq:Veff}, which contains the implicit contribution from the
bare state.\footnote{By using the notations in Sec.~\ref{sec:2-4}, the
  two-body interaction operator $\hat{\mathcal{V}}^{\rm eff} (s)$ can
  be defined as $\hat{\mathcal{V}}^{\rm eff} (s) \equiv
  \hat{\mathcal{V}} + \hat{\mathcal{V}} | \Psi_{0} \rangle \langle
  \Psi_{0} | \hat{\mathcal{V}}/(s-M_{0}^{2})$ in a similar manner to
  the operator $\hat{V}^{\rm eff} (E)$ in Sec.~\ref{sec:2-1}.}  Here
we also assume that the form factor $f(q^{2})$ depends only on the
three momentum so as to make a three-dimensional reduction of the
scattering equation.  Then, by using $\hat{\mathcal{G}}$ and
$\hat{\mathcal{V}}^{\rm eff}$, we can express the wave equation for a
relativistic resonance state $| \Psi )$, whose mass and width are
described by an eigenvalue $s_{\rm R}$, as
\begin{equation}
\hat{\mathcal{G}}^{-1} (s_{\rm R}) | \Psi ) 
= \hat{\mathcal{V}}^{\rm eff} (s_{\rm R}) | \Psi ) ,
\quad 
( \Psi | \hat{\mathcal{G}}^{-1} (s_{\rm R}) 
= ( \Psi | \hat{\mathcal{V}}^{\rm eff} (s_{\rm R}) ,
\label{eqA:Psi_eq}
\end{equation}
which is equivalent to the wave equation in Eq.~\eqref{eq:KVPsi} with
the implicit bare-state degree of freedom.

Let us now derive the scattering equation with the above
normalizations.  To this end, we define the $T$-operator
$\hat{\mathcal{T}}$ by the interaction $\hat{\mathcal{V}}^{\rm eff}$
and two-body Green's operator $\hat{\mathcal{G}}$ as:
\begin{equation}
\hat{\mathcal{T}} = \hat{\mathcal{V}}^{\rm eff} 
+ \hat{\mathcal{V}}^{\rm eff} \hat{\mathcal{G}} \hat{\mathcal{T}} .
\label{eqA:opT}
\end{equation}
This corresponds to the two-body scattering equation in an operator
form.  For the separable interaction~\eqref{eqA:V}, the matrix element
of the $T$-operator is given in the form $\langle
\bm{q}^{\prime \, \text{co}} | \hat{\mathcal{T}} | \bm{q}^{\rm co}
\rangle = T(s) f ( q^{2} ) f ( q^{\prime \, 2} )$.  The scattering
equation is then obtained from Eq.~\eqref{eqA:opT} as
\begin{equation}
T(s) = V^{\rm eff}(s)
+ V^{\rm eff}(s) G(s) T(s) .
\label{eqA:T}
\end{equation}
where $G(s)$ corresponds to the loop function and is defined as 
\begin{align}
  G ( s ) \equiv 
  \int \frac{d^{3} q}{(2 \pi )^{3}} \frac{\sqrt{s_{q}}}{2 \omega
    (\bm{q}) \Omega (\bm{q})} \frac{[f ( q^{2} )]^{2}}{s - s_{q}} 
  = i \int \frac{d^{4} q}{(2 \pi )^{4}} \frac{[f (
    q^{2} )]^{2}} {[(P/2 + q)^{2} - m^{2}] [(P/2 - q)^{2} - M^{2}]} ,
%  \notag \\
\label{eqA:Gloop}
\end{align}
with $P^{\mu} \equiv (\sqrt{s}, \, \bm{0})$.  The second term of the
right-hand side in Eq.~\eqref{eqA:T} can be obtained by inserting the
operator $\hat{\mathcal{P}}_{\rm two} = 1$~\eqref{eqA:unity_two}
between $\hat{\mathcal{V}}^{\rm eff}$ and $\hat{\mathcal{G}}$ as
\begin{align}
\langle \bm{q}^{\prime \, \text{co}} | 
\hat{\mathcal{V}}^{\rm eff} \hat{\mathcal{G}} \hat{\mathcal{T}} 
| \bm{q}^{\rm co} \rangle
= & 
\int \frac{d^{3} q^{\prime \prime}}{(2 \pi )^{3}}
\frac{\sqrt{s_{q^{\prime \prime}}}}{2 \omega (\bm{q}^{\prime \prime})
\Omega (\bm{q}^{\prime \prime})}
\frac{\langle \bm{q}^{\prime \, \text{co}} | 
\hat{\mathcal{V}}^{\rm eff}
| \bm{q}^{\prime \prime \, \text{co}} \rangle 
\langle \bm{q}^{\prime \prime \, \text{co}} | 
\hat{\mathcal{T}} | \bm{q}^{\rm co} \rangle}
{s - s_{q^{\prime \prime}}}
\nonumber \\
= & \int \frac{d^{3} q^{\prime \prime}}{(2 \pi )^{3}}
\frac{\sqrt{s_{q^{\prime \prime}}}}{2 \omega (\bm{q}^{\prime \prime})
\Omega (\bm{q}^{\prime \prime})}
\frac{V^{\rm eff} (s) f ( q^{\prime \prime \, 2} ) f ( q^{\prime \, 2}) 
\times T (s) f ( q^{2} ) f ( q^{\prime \prime \, 2}) }
{s - s_{q^{\prime \prime}}} 
\nonumber \\
= & V^{\rm eff}(s) G(s) T(s) f ( q^{2} ) f ( q^{\prime \, 2}) .
\end{align}
As seen in the last expression of the loop function $G(s)$ in
Eq.~\eqref{eqA:Gloop}, Eq.~\eqref{eqA:T} is nothing but the scattering
equation with the Klein-Gordon propagators, and hence the wave
equation~\eqref{eq:KVPsi} indeed describes a two-body system whose
motion is governed by the Klein-Gordon equation.

At last we emphasize that the normalization~\eqref{eqA:norm_rel} is
consistent with the two-body Green's operator $\hat{\mathcal{G}}(s) =
1 / (s - \hat{\mathcal{K}})$, which is a natural extension of the
nonrelativistic Green's operator $\hat{\mathcal{G}}(E) = 1 / (E -
\hat{H}_{0})$.  Otherwise, we should redefine $\hat{\mathcal{G}}(s)$
so as to absorb a kinematical factor coming from $\sqrt{s_{q}} / [2
\omega (\bm{q}) \Omega (\bm{q})]$ in the loop
integral~\eqref{eqA:Gloop}.  This allows us to determine the
coefficient of the relativistic two-body wave function in
Sec.~\ref{sec:2-4} in a straightforward way as in the nonrelativistic
case.

\end{document}